\documentclass[twocolumn]{aastex631}

\usepackage{graphbox,graphicx}
\usepackage{hyperref} 
\usepackage{booktabs}

\shorttitle{Investigating the intracluster medium viscosity using the tails of GASP jellyfish galaxies}
\shortauthors{Ignesti et al.}

\begin{document}
\title{Investigating the intracluster medium viscosity using the tails of GASP jellyfish galaxies}

\correspondingauthor{Alessandro Ignesti}
\email{alessandro.ignesti@inaf.it}

\author[0000-0003-1581-0092]{Alessandro Ignesti}\affiliation{INAF-Padova Astronomical Observatory, Vicolo dell’Osservatorio 5, I-35122 Padova, Italy}

\author[0000-0003-4195-8613]{Gianfranco Brunetti}\affiliation{INAF, Istituto di Radioastronomia di Bologna, via Piero Gobetti 101, 40129 Bologna, Italy}

\author[0000-0002-7296-9780]{Marco Gullieuszik}\affiliation{INAF-Padova Astronomical Observatory, Vicolo dell’Osservatorio 5, I-35122 Padova, Italy}

\author[0000-0001-7011-9291]{Nina Akerman}\affiliation{INAF-Padova Astronomical Observatory, Vicolo dell’Osservatorio 5, I-35122 Padova, Italy}\affiliation{Dipartimento di Fisica e Astronomia ``Galileo Galilei'', Università di Padova, vicolo dell'Osservatorio 3, IT-35122, Padova, Italy}

\author[0000-0002-5655-6054]{Antonino Marasco}\affiliation{INAF-Padova Astronomical Observatory, Vicolo dell’Osservatorio 5, I-35122 Padova, Italy}

\author[0000-0001-8751-8360]{Bianca M. Poggianti}\affiliation{INAF-Padova Astronomical Observatory, Vicolo dell’Osservatorio 5, I-35122 Padova, Italy}

\author[0000-0001-5262-6150]{Yuan Li}\affiliation{Department of Astronomy, University of Massachusetts, Amherst, MA 01003, USA}

\author[0000-0003-0980-1499]{Benedetta Vulcani}\affiliation{INAF-Padova Astronomical Observatory, Vicolo dell’Osservatorio 5, I-35122 Padova, Italy}

\author[0000-0002-0843-3009]{Myriam Gitti}\affiliation{Dipartimento di Fisica e Astronomia, Università di Bologna, via Piero Gobetti 93/2, 40129 Bologna, Italy}\affiliation{INAF, Istituto di Radioastronomia di Bologna, via Piero Gobetti 101, 40129 Bologna, Italy}

\author[0000-0002-1688-482X]{Alessia Moretti}\affiliation{INAF-Padova Astronomical Observatory, Vicolo dell’Osservatorio 5, I-35122 Padova, Italy}

\author[0000-0002-3818-1746]{Eric Giunchi}\affiliation{Dipartimento di Fisica e Astronomia, Università di Bologna, via Piero Gobetti 93/2, 40129 Bologna, Italy}
\author[0000-0002-8238-9210]{Neven Tomi\v{c}i\'{c}}\affiliation{Department of Physics, Faculty of Science, University of Zagreb, Bijenicka 32, 10 000 Zagreb, Croatia}
\author[0000-0002-8372-3428]{Cecilia Bacchini}\affiliation{DARK, Niels Bohr Institute, University of Copenhagen, Jagtvej 155, 2200 Copenhagen, Denmark}
\author[0000-0001-9143-6026]{Rosita Paladino}\affiliation{INAF, Istituto di Radioastronomia di Bologna, via Piero Gobetti 101, 40129 Bologna, Italy}
\author[0000-0002-3585-866X]{Mario Radovich}\affiliation{INAF-Padova Astronomical Observatory, Vicolo dell’Osservatorio 5, I-35122 Padova, Italy}
\author[0000-0001-5840-9835]{Anna Wolter}\affiliation{INAF - Osservatorio Astronomico di Brera, via Brera, 28, 20121, Milano, Italy}

\begin{abstract}
The microphysics of the intracluster medium (ICM) in galaxy clusters is still poorly understood. Observational evidence suggests that the effective viscosity is suppressed by plasma instabilities that reduce the mean free path of particles. Measuring the effective viscosity of the ICM is crucial to understanding the processes that govern its physics on small scales. The trails of ionized interstellar medium left behind by the so-called jellyfish galaxies can trace the turbulent motions of the surrounding ICM and constrain its local viscosity. We present the results of a systematic analysis of the velocity structure function (VSF) of the H$\alpha$ line for ten galaxies from the GASP sample. The VSFs show a sub-linear power law scaling below 10 kpc which may result from turbulent cascading and extends to 1 kpc, below the supposed ICM dissipation scales of tens of kpc expected in a fluid described by Coulomb collisions. Our result constrains the local ICM viscosity to be 0.3-25$\%$  of the expected Spitzer value. 
Our findings demonstrate that either the ICM particles have a smaller mean free path than expected in a regime defined by Coulomb collisions, or that we are probing effects due to collisionless physics in the ICM turbulence.
\end{abstract}


\section{Introduction} \label{sec:intro}
The intracluster medium (ICM) microphysics in galaxy clusters is poorly known. It is usually assumed to be a fluid dominated by Coulomb collisions, which can explain its large-scale properties  \citep[e.g.,][]{Sarazin_1988}. However, the properties of the gas density fluctuations \citep[][]{Zhuravleva_2019}, and the observational evidence of cosmic ray electron re-accelerated by ICM turbulence \citep[e.g.,][]{Brunetti-Jones_2014}, indicate that the effective ICM viscosity is orders of magnitude smaller than expected from the thermal ion-ion Coulomb collisions \citep[e.g.,][]{Spitzer_1962}. This is because the ICM is a `weakly collisional' plasma due to various plasma instabilities, \citep[e.g.,][for a review]{Schekochihin_2022} that can perturb the ICM magnetic field on small scales, thus strongly reducing the effective ICM particles' mean free path, and, hence, also the effective viscosity of the fluid.

Cluster galaxies can be used to probe the ICM properties, especially the so-called jellyfish galaxies \citep[e.g.,][]{Smith2010,Ebeling2014,Fumagalli2014,Poggianti2017,Boselli_2022}, that are the result of infalling spiral galaxies interacting with the ICM. These galaxies show characteristic `tails' of ionized plasma, produced by the interstellar medium (ISM) being displaced outside of the stellar disk via ram pressure stripping \citep[RPS,][]{Gunn1972}, where the ISM and the ICM can interact via mixing. Observational evidence of ISM-ICM mixing has been collected in various forms, from metallicity gradients along the tails \citep[][]{Franchetto_2021}, to extended X-ray emission associated with the H$\alpha$ emission \citep[e.g.,][]{Sun2010, Poggianti_2019,Campitiello_2021,Sun_2022,Bartolini_2022}, to peculiar optical line ratios \citep[][]{Poggianti2019,Campitiello_2021}, to the presence of diffuse ionized gas \citep[][]{Tomicic_2021,Pedrini_2022}, or large-scale magnetic fields accreted from the ICM \citep[][]{Muller_2021}. All these results point toward the fact that, in the tails of jellyfish galaxies, the stripped ISM dynamic is affected by the ICM small-scale motions, such as turbulence. Therefore, the diffuse ISM H$\alpha$ emission might probe the ICM turbulent motions and, thus, trace the turbulent cascade down to its dissipation scale, hence constraining the ICM viscosity. \citet[][]{Li_2023} pioneered this method by analyzing the velocity structure function (VSF) in the tail of the nearby jellyfish galaxy ESO137-001. The VSF is a two-point correlation function that quantifies the kinetic energy fluctuations as a function of the scale $l$ in a velocity field. The VSF can trace both the energy cascade expected in turbulent flows, and coherent motions such as collapse, rotation, or blast waves, which appear in the form of coherent velocity differences \citep[e.g.,][]{Kolmogorov_1941,Heyer_2004,Chira_2019}. For fluid particles in fully developed, homogeneous, and isotropic turbulence, \citet[][]{Kolmogorov_1941} predicted that the VSF is well-described by a power-law relation which extends down to a dissipation scale, $\eta$, which depends on the fluid viscosity. \citet[][]{Li_2023} observed a turbulent cascade in the VSF of ESO137-001 extending for two orders of magnitude below the supposed dissipation scale, thus constraining the ICM viscosity to $\sim$0.01 of the predicted value. 

In this paper, we extend the VSF analysis to a sample of ten galaxies from the GASP survey\footnote{\url{https://web.oapd.inaf.it/gasp/}} \citep[GAs Stripping Phenomena in galaxies with MUSE,][]{Poggianti2017}. By operating on a larger sample, we can simultaneously test the results of \citet[][]{Li_2023}, and explore the variation in ICM viscosity for different galaxy cluster properties and regions. The manuscript is structured as follows. In Section \ref{sec_data} we present the sample and the data preparation, from MUSE data processing to the VSF construction. In Section \ref{sec_dis} we report and discuss the results, and in Section \ref{sec_caveats} we list the main caveats of our analysis. Finally, in Appendix \ref{app_rot_corr} and \ref{app_models} we report a series of diagnostic plots, in Appendix \ref{appendix_vsf} we show the VSF for the entire sample, and in Appendix \ref{simu} we describe the simulation setup adopted to compare the observed VSFs. In this work we adopt the standard concordance 
cosmology parameters $H_0 = 70 \, \rm km \, s^{-1} \, Mpc^{-1}$, ${\Omega}_M=0.3$
and ${\Omega}_{\Lambda}=0.7$. At the redshifts of the clusters in which galaxies are located,  $1''\simeq0.9-1.1$ kpc.
\section{Data analysis}
\label{sec_data}
\subsection{Sample selection}
\begin{figure}
    \centering
    \includegraphics[width=.5\textwidth]{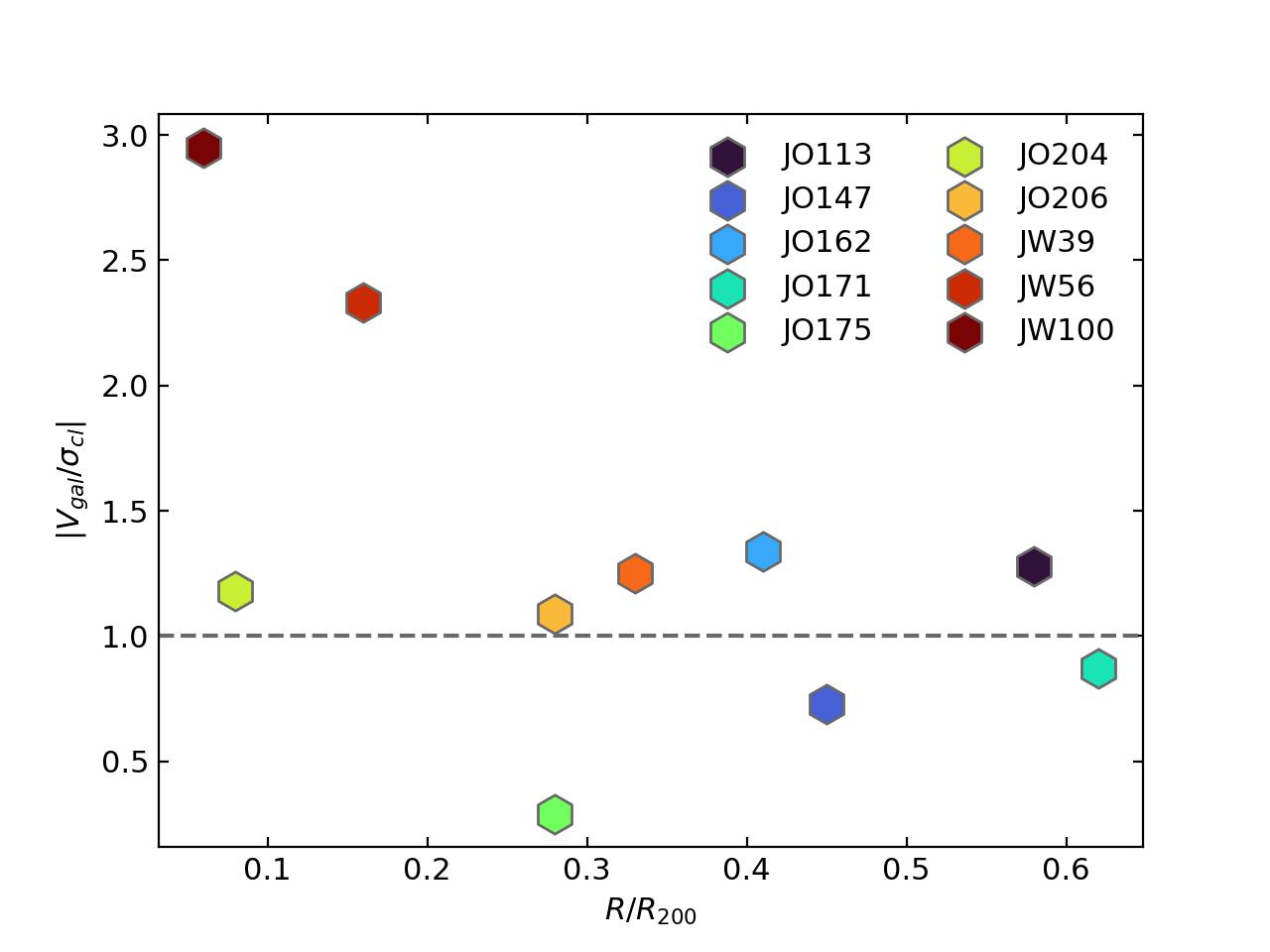}
    
    \caption{Line-of-sight velocity in units of the hosting cluster velocity dispersion, vs. projected position, in units of $R_{200}$. The horizontal dashed line indicates $|V_{\text{gal}}|=\sigma_{\text{cl}}$.}
    \label{img_phase_space}
\end{figure}

To investigate the properties of the stripped ISM, we select the most extreme ram pressure stripped galaxies from the GASP sample, which is composed of galaxies at $z$=0.04-0.07 located in different environments, from the field to clusters. Specifically, we select the ten galaxies in clusters with the most extended H$\alpha$ tails in the plane of the sky in the MUSE images. Their properties, and those of their hosting clusters, which are presented in \citet[][]{Poggianti2017,Gullieuszik_2020}, are summarised in Table \ref{table_sample}. The resulting sample spans more than two orders of magnitude in stellar mass, from log$(M_{\text{star}}/M_\odot)=9.05$ (JW56) to log$(M_{\text{star}}/M_\odot)=11.47$ (JW100) and covers different regions in the phase-space diagram (Figure \ref{img_phase_space}), which indicates that galaxies are in different ICM and RPS conditions.

These galaxies, and their stripped tails, have been the subject of previous studies. They all show H$\alpha$ tails with projected lengths of a few tens of kpc that host extraplanar star formation \citep[][]{Poggianti2019,Tomicic_2021}. In the MUSE spectra, JW56 and JW100 show extended regions in which the optical line ratios are consistent with LINERS emission \citep[][]{Poggianti2019}, indicating that the main ionization mechanism could be the photo-ionization due to the ICM thermal emission \citep[][]{Campitiello_2021}. For JW100, \textit{Chandra} observations detected extended X-ray emission in the tail, which correlates spatially with the H$\alpha$ emission \citep[][]{Poggianti_2019} and supports the hypothesis of ongoing ISM-ICM mixing \citep[][]{Sun_2022}. Finally, in JO206, JW39 and JW100 show non-thermal radio tails \citep[][]{Muller_2021, Ignesti_2022d}, which is evidence of large-scale magnetic fields extending in the tails.

\subsection{MUSE data preparation}
\label{data_analysis}

This work is based on the kinematics of the ionized gas derived from the H$\alpha$ emission.
We use the MUSE data from the GASP survey; the observations, data reduction, and analysis are presented in \citet{Poggianti2017}.
The spatial resolution of these seeing-limited observations is $\sim 1\arcsec$, which at the redshift of the targets corresponds to $\sim1$\,kpc. Emission line fluxes and gas kinematics were measured using the IDL software KUBEVIZ \citep{Fossati2016} on the datacubes obtained by subtracting the stellar component from the observed data and corrected for dust extinction \citep[for details see][]{Poggianti2017}. KUBEVIZ also provides the 1$\sigma$ error for each fitted parameter, including the line of sight velocity. We report the H$\alpha$ line velocity maps in the galaxy rest-frame by normalizing them for the average velocity measured at the center of the stellar disk.

\begin{table*}[]
    \centering
 
    \begin{tabular}{lccccccccc}
 
    \midrule
    \midrule
    Galaxy&$z_{\text{gal}}$&log $M_{\text{star}}/M_\odot$&Cluster&$z_{\text{cl}}$&log $M_{200}/M_{\odot}$&$R_{200}$ [Mpc]&$\sigma_{\text{cl}}$ [km s$^{-1}$]&$R_{\text{gal}}/R_{200}$&$|V_{\text{los}}/\sigma|$\\
    
    \midrule
JO113&0.0552&9.69&A3158&0.05947&14.94&1.94&948&0.58&1.28\\
JO147&0.0506&11.03&A3558&0.04829&14.95&1.95&910&0.45&0.73\\
JO162&0.0454&9.42&A3560&0.04917&14.83&1.79&799&0.41&1.34\\
JO171&0.0521&10.61&A3667&0.05528&15.12&2.22&1031&0.62&0.87\\
JO175&0.0467&10.50&A3716&0.04599&14.78&1.72&753&0.28&0.29\\
JO204&0.0424&10.50&A957&0.04496&14.53&1.42&631&0.08&1.18\\
JO206&0.0511&10.96&IIZW108&0.04889&14.31&1.20&575&0.28&1.09\\
JW39&0.0663&11.21&A1668&0.0634&14.48&1.35&654&0.33&1.25\\
JW56&0.0387&9.05&A1736&0.0461&14.92&1.92&918&0.16&2.33\\
JW100&0.0619&11.47&A2626&0.05509&14.59&1.48&650&0.06&2.95\\

    \midrule
    \end{tabular}
    \caption{Sample properties. From left to right: GASP name; galaxy redshift; Stellar mass; hosting cluster name, redshift, $M_{200}$, $R_{200}$, and velocity dispersion $\sigma_{\text{cl}}$ \citep[][]{Vulcani2018,Gullieuszik_2020}; galaxy clustercentric distance in units of  $R_{200}$; galaxy velocity along the line-of-sight in units of $\sigma_{\text{cl}}$.}
    \label{table_sample}
\end{table*}
\subsection{Velocity structure function}
\label{sec_res}
The stripped tails are complex, filamentary structures in which the properties of the plasma are affected by the presence of star-forming complexes \citep[][]{Poggianti2019,Giunchi_2023}, and by other large-scales dynamical processes which imprint coherent velocity differences in the VSF. Thus it is necessary to further process the data to recover the turbulence signature in the H$\alpha$ VSF.
\subsubsection{Step I: Locating the diffuse ISM}
We first filter the data to extract only the signal from the extraplanar diffuse emission. Specifically, the spaxels of the tails from which to measure the line velocity are selected by applying a series of filters in the H$\alpha$ images. To begin with, we consider only those pixels with a signal-to-noise ratio higher than 5 in the H$\alpha$ flux density, with a flux density lower than $10^{-16}$ erg s$^{-1}$ cm$^{-2}$, to avoid hot pixels, and with an uncertainty on the velocity fit lower than 30 km s$^{-1}$. Moreover, we mask the pixels located inside of the stellar disk \citep[as defined in][]{Gullieuszik_2020} and the star-forming clumps detected in H$\alpha$ \citep[][]{Poggianti2017b}. This latter condition is physically motivated by the consideration that the plasma in the clumps, which surrounds the new stars, is less likely to be bound to the turbulent ICM. This selection provides us with the signal from the extraplanar, diffuse plasma in the tails. From the remaining signal, we further remove the isolated regions with an area smaller than 200 pixels for which it is difficult to ascertain that they really belong to the stripped tails. 
\subsubsection{Step II: Filtering out rotation and advection}
\label{novel}
The stripped ISM dynamic outside the stellar disk is primarily set by the combination of 1) advection induced by the ram pressure, 2) rotation inherited from the disk, and 3) ISM small-scale motions \citep[e.g.,][]{Gullieuszik2017,Tonnesen_2021, Luo_2023, Akerman_2023,Sparre_2024}. To study the ISM small-scale motions and, hence, the ICM turbulence signature in the VSF, it is crucial to remove the other two factors. However, their removal is complicated by the intrinsic filamentary structure of the stripped ISM, which makes it difficult to define an unambiguous rotation axis to conduct an analysis of the 3D rotation \citep[e.g.,][for the analysis of the rotating molecular gas in the disks of the same GASP galaxies]{Bacchini_2023}, and by projection effects, which can lead gas originating from different sides of the disk to overlap along the line of sight. Therefore, we propose a novel technique to remove these components from the velocity field, which we show here for JO206, and for the other galaxies in Appendix \ref{app_rot_corr} and \ref{app_models}.

\begin{figure}
    \centering
    \includegraphics[width=\linewidth]{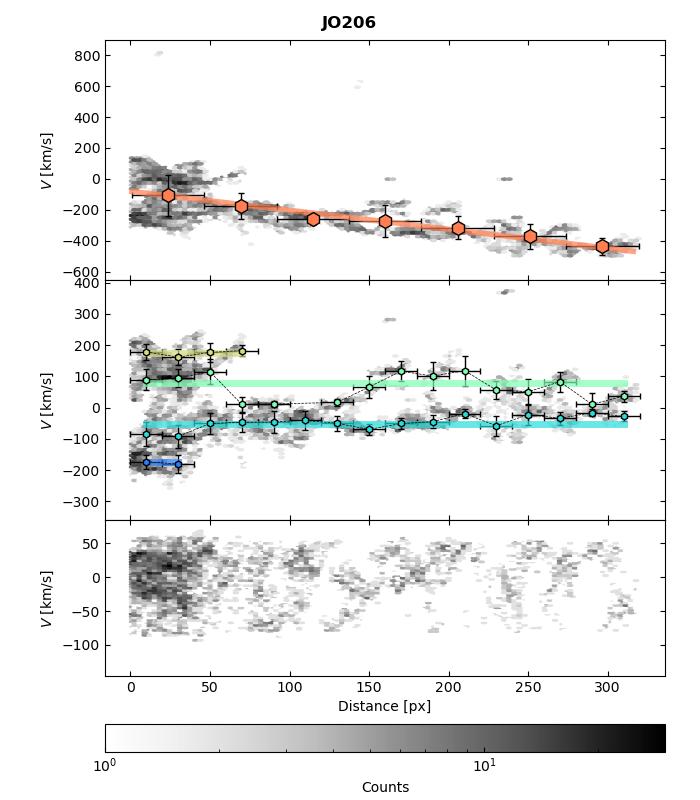}
    \caption{Phase-space density plots derived from the H$\alpha$ velocity map of JO206. Top: initial velocity distribution with the median velocities (orange points) and the best-fitting velocity gradient; Central: Velocity distribution corrected for the systematic velocity. The colored points trace the rotating filaments, and the lines point out the corresponding average velocity; Bottom: Resulting velocity distribution corrected for the rotation of each filament. }
    \label{vel_grad}
\end{figure}

To visualize the structure of the velocity maps, we compose the phase-space diagram for each galaxy, where we compare the velocity measured in each pixel with its minimum distance from the stellar disk edge, as defined in \citep[][]{Gullieuszik_2020}. The diagrams are then converted into density plots to highlight the substructures in the velocity maps (Figure \ref{vel_grad}, top panel). First, we correct the velocity field for the advection component along the line of sight. The ram pressure can induce a velocity increase with the distance from the stellar disk, as the gas clouds gradually approach the ICM wind speed \citep[][]{Luo_2023, Serra_2023}.  So we compute the median velocities at increasing distances from the disk, and use them to fit a linear model of the velocity gradient along the tail (Figure \ref{vel_grad}, top panel, orange points). Then we subtract this systematic component from each pixel depending on their distance from the disk. A similar procedure is presented also by \citet[][]{Luo_2023} to correct the systematic velocity in ESO137-001. This correction can also remove the systematic velocity shift derived from the galaxy velocity in the cluster. The result is a velocity distribution typically centered around zero without clear gradients with the distance (Figure \ref{vel_grad}, central panel). However, we recognize the presence of clear filamentary patterns in the residual phase-space intensity distribution. We assume that these structures are produced by the rotation of the stripped gas inherited from the disc kinematics.

\begin{figure}
    \centering
    \includegraphics[width=\linewidth]{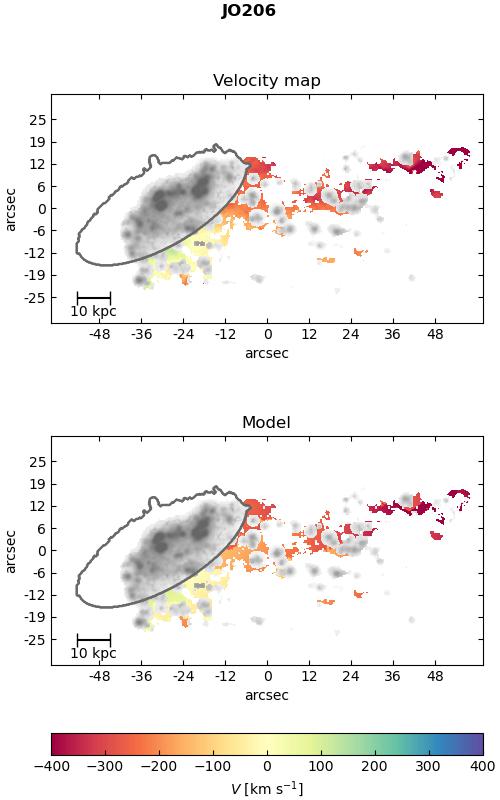}
    \caption{H$\alpha$ observed (top) and rotation+advection model (bottom) velocity map of JO206. We also show the original H$\alpha$ emission (grey colormap), and the stellar disk \citep[grey contour][]{Gullieuszik_2020}. }
    \label{rot_model}
\end{figure}

To remove the rotation, we first identify the different filaments in the phase-space density plots as regions where the points are clustered around certain velocity values. To do so, we bin the phase-space plots in several velocity bins, and then for each bin, we compute the median velocity values at increasing distances from the disk. The numbers of velocity and distance bins are different for each galaxy and they are set by visual inspection. For each filament (colored points in Figure \ref{vel_grad}, central panel), we compute the average velocity, weighted by the number of points in each bin (colored lines in Figure \ref{vel_grad}, central panel). Then we correct each filament by subtracting the corresponding average velocity from each pixel in their bins. At this step, we mask out those pixels located outside the filaments. This operation independently corrects each filament for its velocity, resulting in a velocity distribution centered around zero and without any clear symmetric structure (Figure \ref{vel_grad}, bottom panel). Finally, we further mask out the strong outliers, which typically are residuals left by rotation correction or contaminations from the [NII] line. In Figure \ref{rot_model} we show the original velocity map of the diffuse, stripped ISM (top panel) and the model obtained by combining the advection and rotation components (bottom panel).

\subsubsection{Step III: Computing the VSF}
\begin{figure*}[htb]
\includegraphics[width=\linewidth]{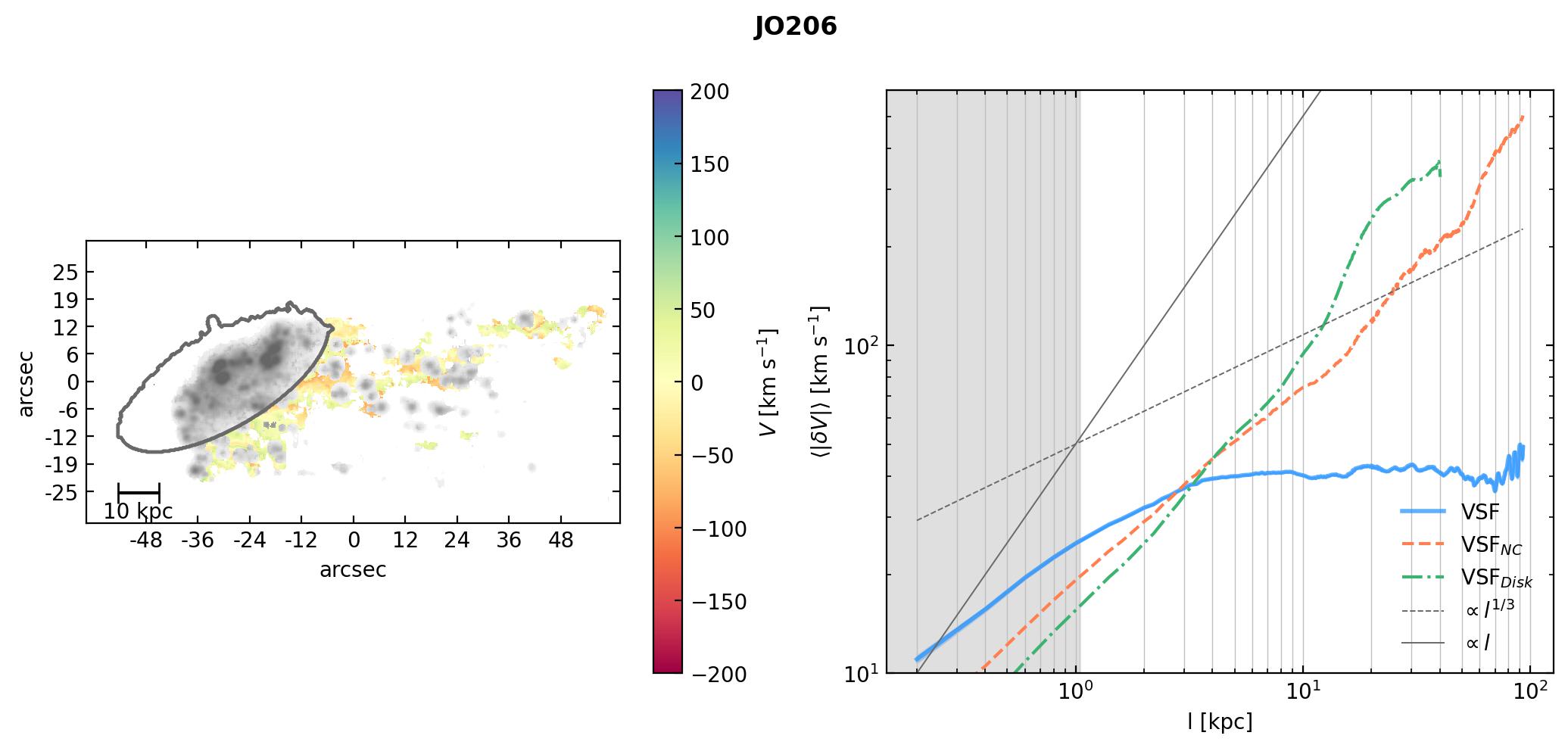}
\caption{\label{img_jo147}Left: Residual velocity map of JO206 (blue-to-red colormap). We also show the original H$\alpha$ emission (grey colormap), and the stellar disk \citep[grey contour][]{Gullieuszik_2020}. Right: VSF of the residual velocity map (blue, the blue-filled area indicates the 1-$\sigma$ uncertainty region), the one derived from the not-corrected one (orange), and the one derived in the stellar disk only (green). The grey-shaded area marks $l<1$ arcsec below which the signal is affected by the spaxel correlation (See Section \ref{data_analysis}). The continuous and dashed lines indicate the slopes expected in the case of, respectively, rotation ($\propto l$) and turbulence ($\propto l^{1/3}$).}
\end{figure*}

The VSFs are computed from the pixels of the residual velocity map as follows: for each pair of pixels, we measure their projected physical distance $l$, which we derive by multiplying their angular separation by the kpc-to-arcsec scale at the hosting cluster redshift, and the velocity difference $\delta V$. The VSF is derived by computing the average absolute value of the velocity differences $\langle |\delta V| \rangle$ within bins of $l$ from 0.2 to 100 kpc with a step of 0.2 kpc, which is approximately the physical scale corresponding to the pixel size of the MUSE images. Correspondingly, by propagating the uncertainties on the velocity measurement of each spaxel, the uncertainties on the fit of the systematic velocity, and the rotating filaments, we compute the error on each bin of the VSF that results in the order of a few km s$^{-1}$. We show in Figure \ref{img_jo147} the residual velocity map (left) and the corresponding VSF (blue line). For comparison, we use the original velocity map not corrected for the rotation and advection to compute the VSF in the tail (VSF$_{\text{NC}}$, orange dashed line), and inside the stellar disk (VSF$_{\text{Disk}}$, green dash-dotted line). The VSFs show different slopes, with the first one resembling the slope $l^{1/3}$ (dashed grey line), and the other ones being generally steeper and more similar to the linear scaling (black line). We note that, as a consequence of atmospheric seeing, the effective angular resolution is limited to $\sim1$ arcsec ($\simeq1$ kpc marked in grey in Figure \ref{img_jo147}), and, thus, below that scale, the VSF signal is less reliable because the pixels are correlated within the resolution elements. Therefore, in the following analysis, we conservatively consider $l=1$ kpc as the lower limits of the VSFs. As a caveat, it is important to remember that this analysis is subject to projection effects as filaments that are physically distant can appear adjacent along the line-of-sight. We refer to Section \ref{sec_caveats} for a detailed discussion of the role of projection effects.

\subsection{Comparison with a simulated case}

\begin{figure}
    \centering
    
    \includegraphics[width=\linewidth]{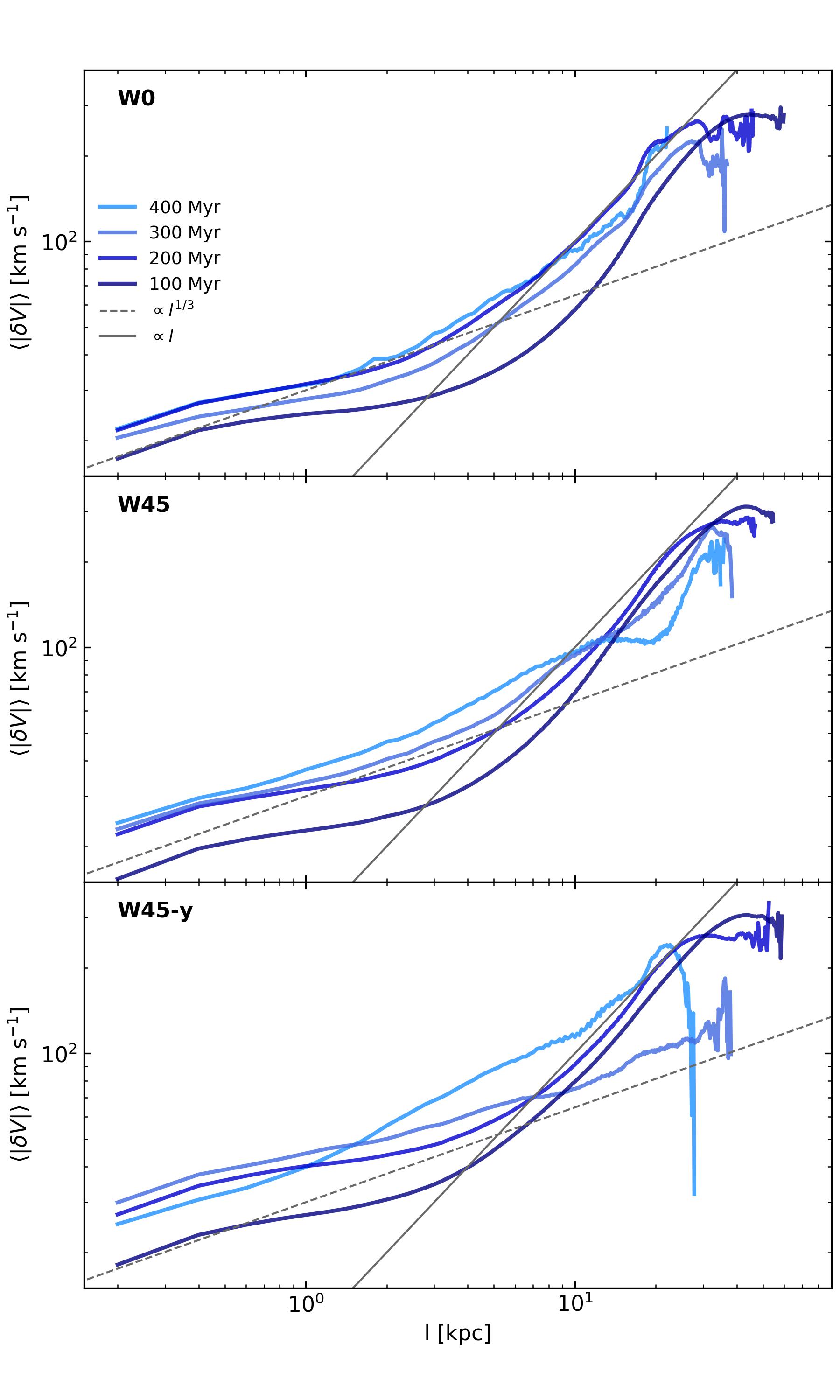}
    \caption{VSFs of the simulated tail for different inclinations (top to bottom) at different snapshots. The VSFs reported here are derived without prior reprocessing of the simulated velocity maps. The continuous and dashed lines indicate the slopes expected in the case of, respectively, rotation ($\propto l$) and turbulence ($\propto l^{1/3}$). }
    \label{fig_test_VSF}
\end{figure}

To further corroborate our analysis, we explore the case of the VSF in the absence of ICM turbulence. To do so, we make use of the simulation setup presented in \citet[][]{Akerman_2023} (see Appendix \ref{simu}). The simulations model stripping of a massive Milky-Way-like galaxy with a radius of $\sim30$ kpc and which rotates at $\sim200$ km/s. In this set-up, the gas dynamics are set solely by rotation and ram pressure and the simulated ICM wind is initially laminar at the galaxy scale. Furthermore, there are no magnetic fields. We produced maps of the gas velocity along the line-of-sight for the gas phase in the temperature interval $3.5<$logT$<6.5$, each one at four different snapshots in time (100, 200, 300, and 400 Myr after the beginning of the stripping). The maps capture the gas in the stripped tails, defined as 2 kpc above the galaxy plane in the direction of stripping. Specifically, we analyze the case of face-on stripping (W0) and inclined stripping (W45, wind hits the galaxy at $45^\circ$). In W0 the wind blows along the $z$-axis, while in W45 the wind has components along both the $y$- and the $z$-axes. The latter ensures that the line-of-sight is aligned with the stripping direction and allows us to further investigate the role of projection effects.
For the simulated VSFs, we take the gas projections along the $x$-axis for both W0 and W45 galaxies and an additional projection along the $y$-axis for W45 (`W45-y' in Figure \ref{fig_test_VSF}). 
We set the pixel size of the velocity maps to 0.2 kpc to match the physical scale of the MUSE images. The VSFs are computed with the same method adopted for the real MUSE observations, that is by considering only the gas outside of the stellar disk. We note that, at this stage, we do not attempt to remove the rotation and advection. The resulting VSFs are shown in Figure \ref{fig_test_VSF}, divided by simulation run (rows). The velocity maps themselves can be found in Appendix \ref{simu}.

In the early stages (100-200 Myr), when the tail is still rich in rotating gas (see Figure \ref{im_simu}), the VSFs show a recurrent shape characterized by an almost constant slope from 0.2 to 2-3 kpc, a strong inflection between 3 and 7 kpc, and linear slope expected by the rotation up to the disk radius scale. As a cautionary tale, we note that for the late-stage stripping, when the tail has already lost most of the gas and what is left does not keep track of the disk rotation anymore the VSFs are flatter than the initial-stage stripping case. This result suggests that an advection-dominated tail, in which the gas particles move with a uniform velocity set by the ram pressure wind, results in a sub-linear VSF, which is similar to a turbulent-dominated one.

\begin{figure*}
\includegraphics[width=0.33\textwidth,align=c]{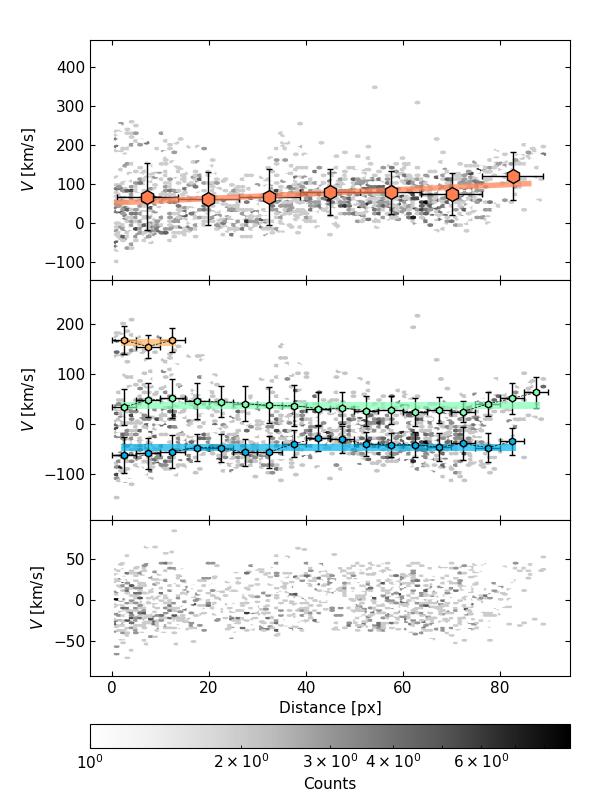}
\includegraphics[width=0.33\textwidth,align=c]{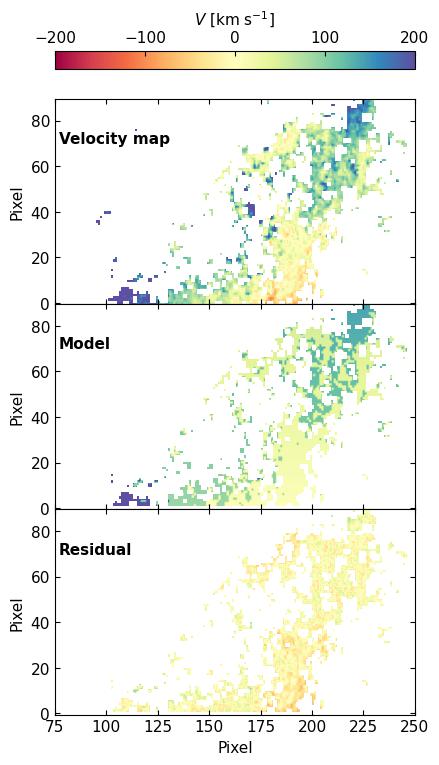}
\includegraphics[width=0.33\textwidth,align=c]{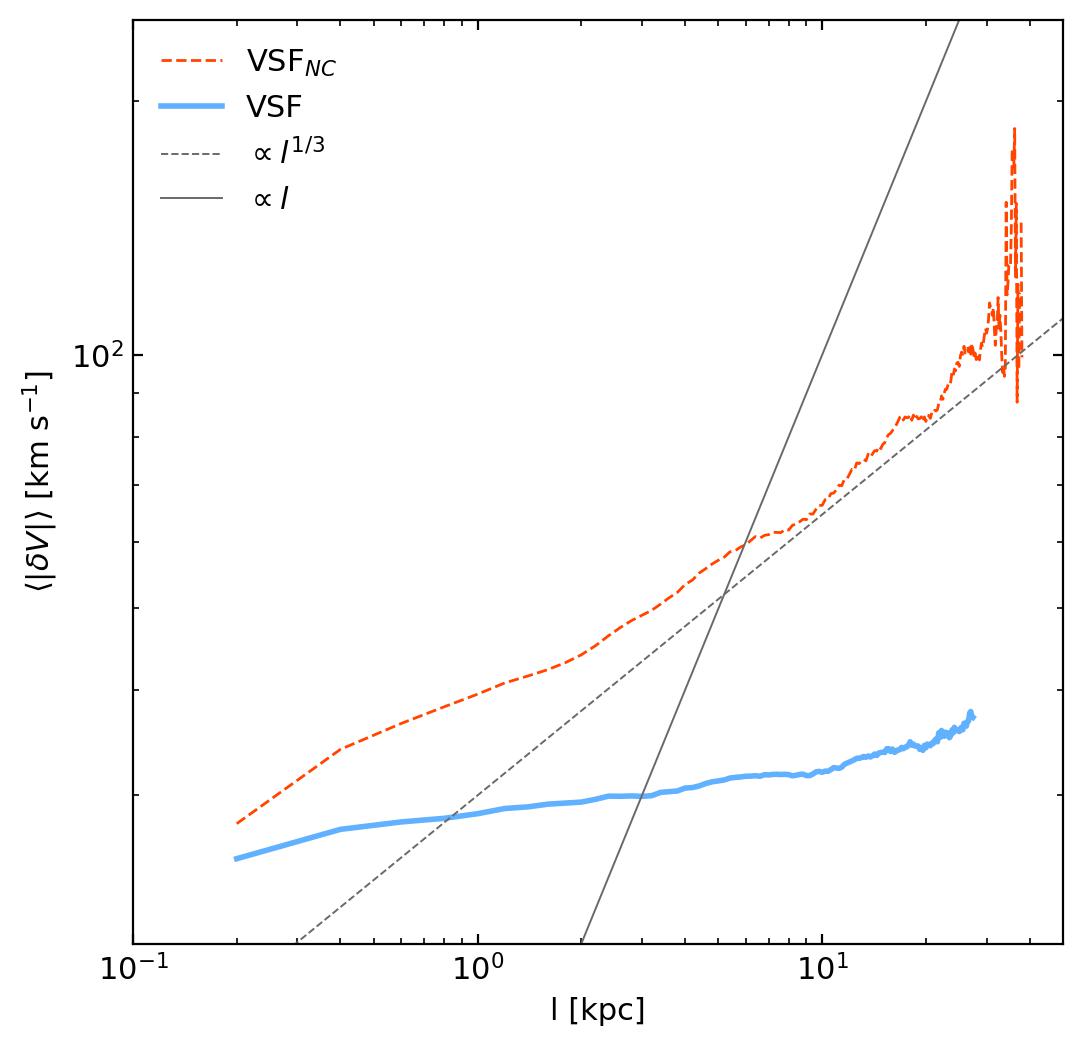}
\caption{\label{sium} Correction procedure applied on the simulated tail W45-y at 300 Myr after the stripping. Left: Phase-space density plot for the simulated tail; Middle: Comparison between the original (top), model (center), and residual (bottom) velocity maps; Right: VSF before (orange) and after (blue) the correction procedure (Section \ref{novel}). The continuous and dashed lines indicate the slopes expected in the case of, respectively, rotation ($\propto l$) and turbulence ($\propto l^{1/3}$).}
\end{figure*}

Finally, we use the simulated tails to demonstrate the effectiveness of the correction procedure described in Section \ref{novel}. Specifically, we perform the procedure on the W45-y run, at 300 Myr after the beginning of the stripping, because of the complex morphology composed of filamentary structures visually resembling the observed tails. We show each step and the final results in Figure \ref{sium}. After the removal of rotation and advection, the residual VSF (blue line) results significantly flattened, which denotes the fact that the residual velocity map is devoid of large-scale coherent velocity differences, thus confirming the efficacy of the procedure. We note that the residual VSF is also flatter than the turbulent slope, which indicates that the simulated tails lack small-scale turbulent motions. We conclude that it is likely due to the limits of the simulation setup (see Appendix \ref{simu}) which cannot fully resolve the turbulence development. Although done intentionally, as such a configuration is designed to save on computational time and resources and, furthermore, the simulation was designed for a different scientific analysis \citep[see][]{Akerman_2023}, this means that future studies aiming to focus on the development of stripped tails will require tailored numerical simulations.

\section{Results and Discussion}
\begin{figure}
    
    \includegraphics[width=.5\textwidth]{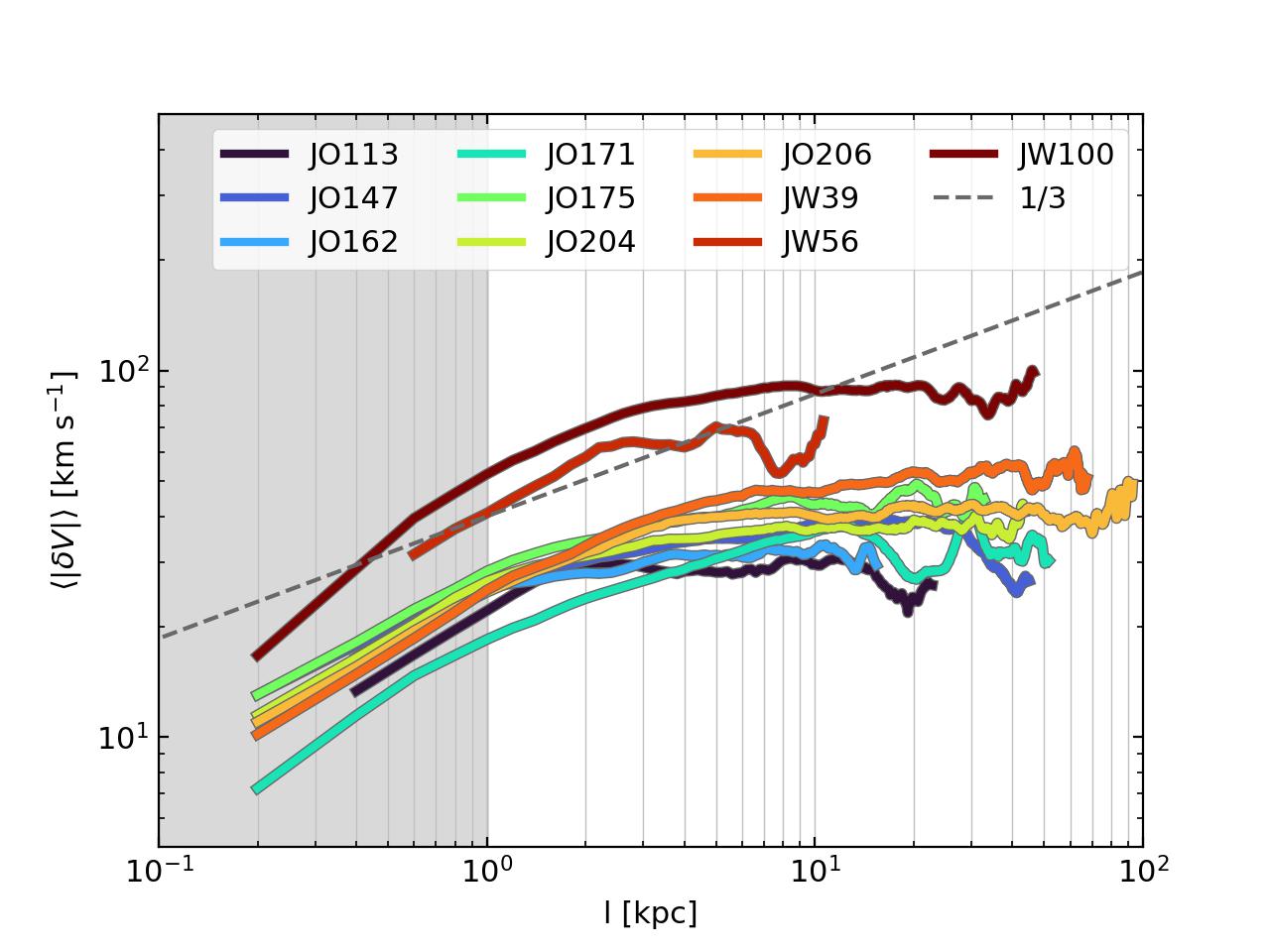}
   \caption{Comparison of the VSF of the entire sample. The black-dashed line indicates the $l^{1/3}$ trend. The grey-filled region marks $l<1$ arcsec below which the signal is affected by the spaxel correlation (See Section \ref{data_analysis}). }
    \label{img_VSF}
\end{figure}

\label{sec_dis}
\subsection{Results}
In Figure \ref{img_VSF} we show the resulting, corrected VSF for each galaxy compared with the Kolmogorov spectrum $\langle |\delta V| \rangle\propto l^{1/3}$ \citep[][]{Kolmogorov_1941}. The individual VSFs are presented in Appendix \ref{appendix_vsf}. Generally, the galaxies approximately follow the $\propto l^{1/3}$ scaling below 5 kpc, whereas they are flat on larger scales due to the removal of rotation and advection, which dominate the large-scale dynamic. This pattern is consistent with the one shown by \citet[][]{Li_2023}. 

The residual VFSs are different from the simulated ones (Figure \ref{fig_test_VSF}), lacking both the steep, high-velocity part at large scales, induced by the combination of rotation and advection, and the flattening at small scales. On the contrary, these features are present in the not-corrected VSFs (Appendix \ref{appendix_vsf}, orange lines). Therefore, we argue that we have removed most of the rotation and advection components, and thus, the residual VSFs trace the small-scale, turbulent motions of the stripped ISM.

\subsection{What drives the turbulence in the tails?}
\label{3.2}
The turbulent cascade we observe could either result from the `frozen-in' ISM turbulence inherited from the disk or the ICM turbulence. To test the first scenario, in Figures \ref{img_jo147} and \ref{app_vsf} we compare the VSF computed inside (green lines) and outside (orange lines) of the stellar disk. We observe two cases, galaxies in which the disk VSF is lower than the tail one (JO113, JO162, JO175, JO206, and JW39), and galaxies in which we observe the opposite (JO147, JO171, JO204, JW56, and JW100). The first case naturally challenges the frozen-in ISM turbulence scenario because it indicates that the ISM turbulence can be weaker, and the tail turbulence is likely dominated by the growth of instabilities induced by the interaction with the ICM. Regarding the second case, we observe that the ISM turbulence can be significantly strong (e.g., the case of JW56 and JW100), but once the gas is stripped, it decays quickly. For JW56 and JW100, the flattening of the VSF takes place at the 2-3 kpc scale with $\delta v\sim70$ km s$^{-1}$. The driving scale is typically a factor 2-3 larger than the flattening scale \citep[][]{Zhang_2024}, thus we estimate an eddy turnover time ($t\simeq l/v$) of $\sim50-100$ Myr. Assuming an average stripped plasma velocity of 300 km s$^{-1}$, which is in line with the constraints provided by the decay of the radio continuum emission in ram pressure stripped tails \citep[][]{Ignesti_2023, Roberts_2024}, then within one eddy turnover time, the stripped ISM has moved about $\sim15-30$ kpc. This length scale is consistent with the observed tail length. So we argue that what we observe in the tail is never pure, frozen-in ISM turbulence but we are tracing fully mixed ISM or, at least, stripped ISM in a transition stage.

\begin{figure*}[t]
\includegraphics[width=.5\textwidth]{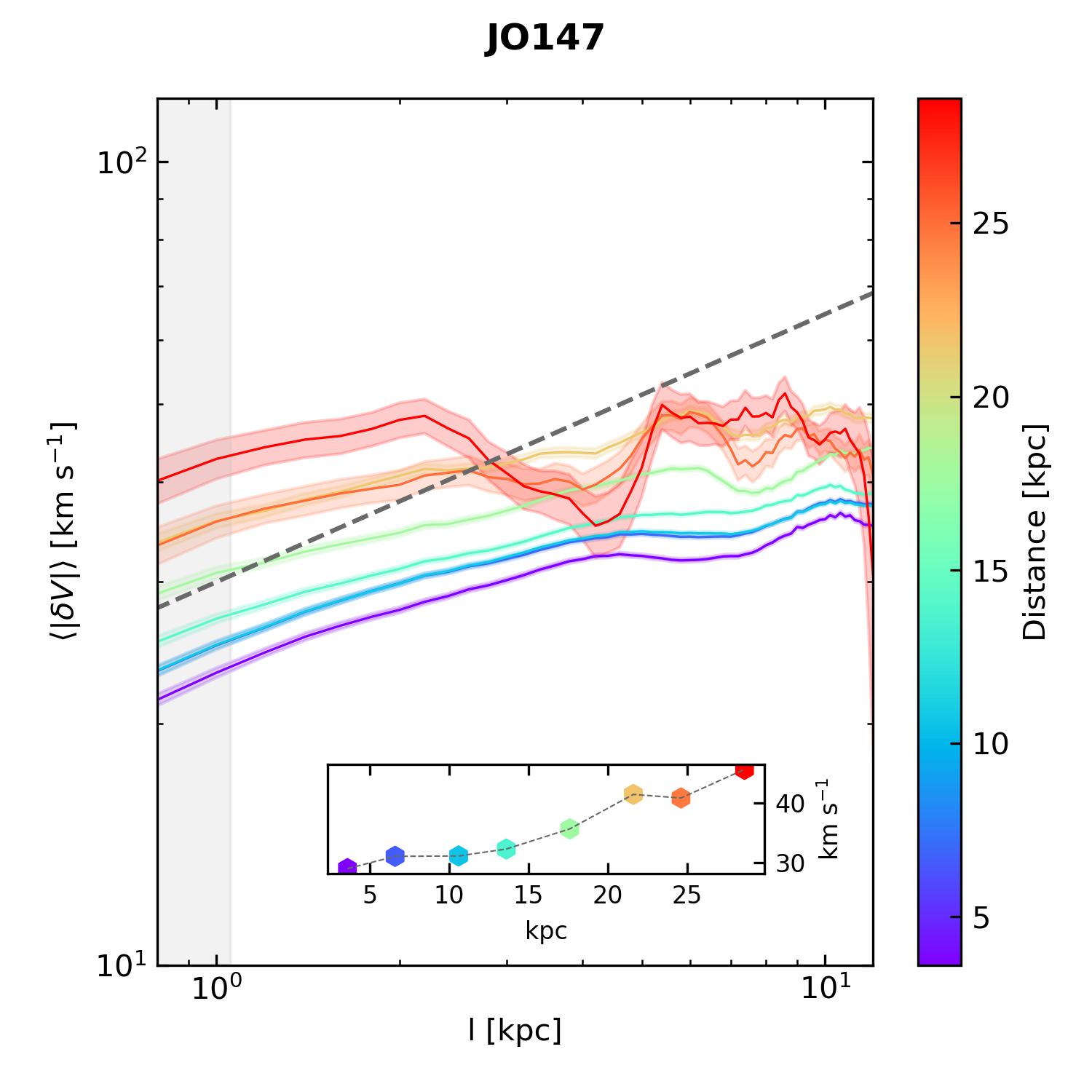}
\includegraphics[width=.5\textwidth]{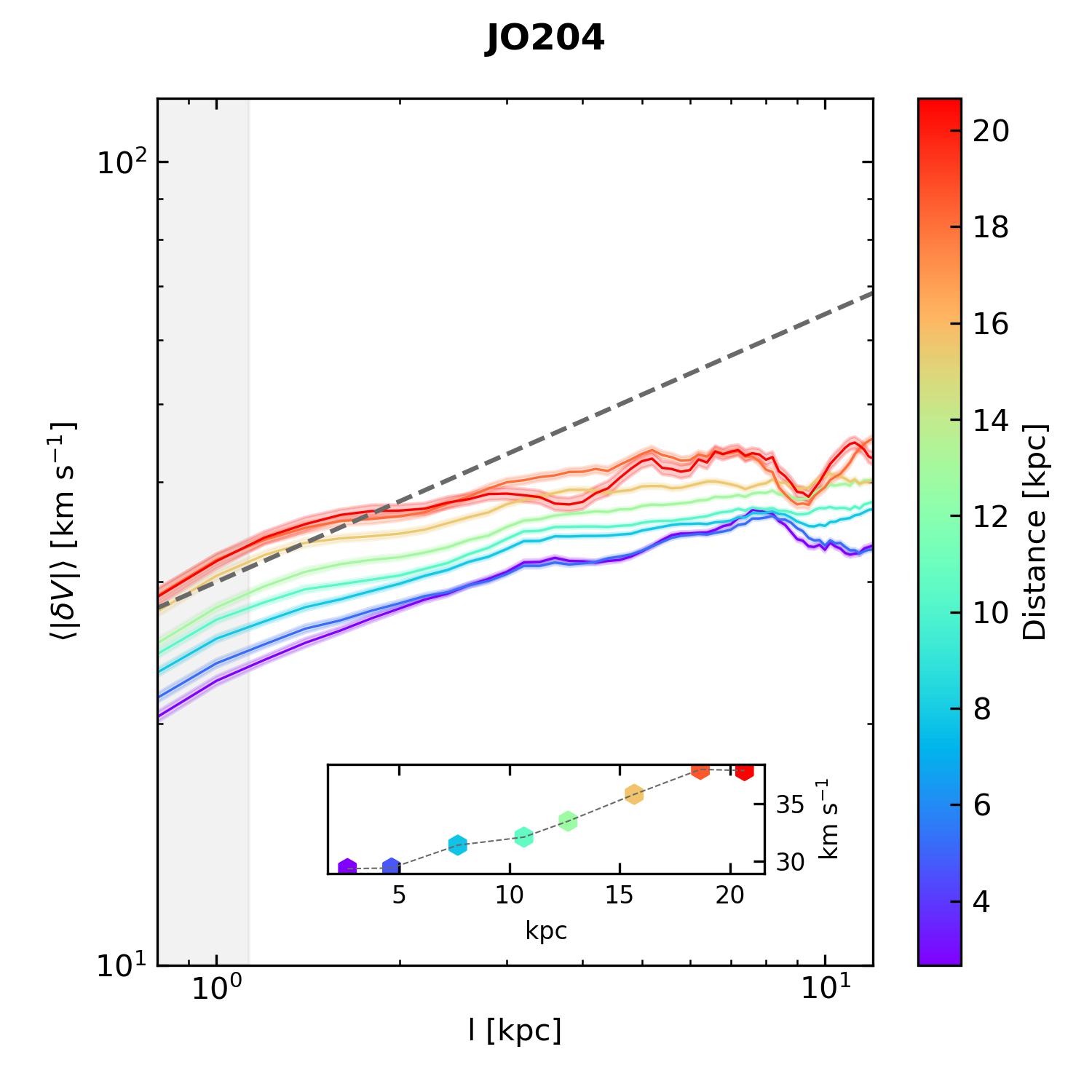}\\
\includegraphics[width=.5\textwidth]{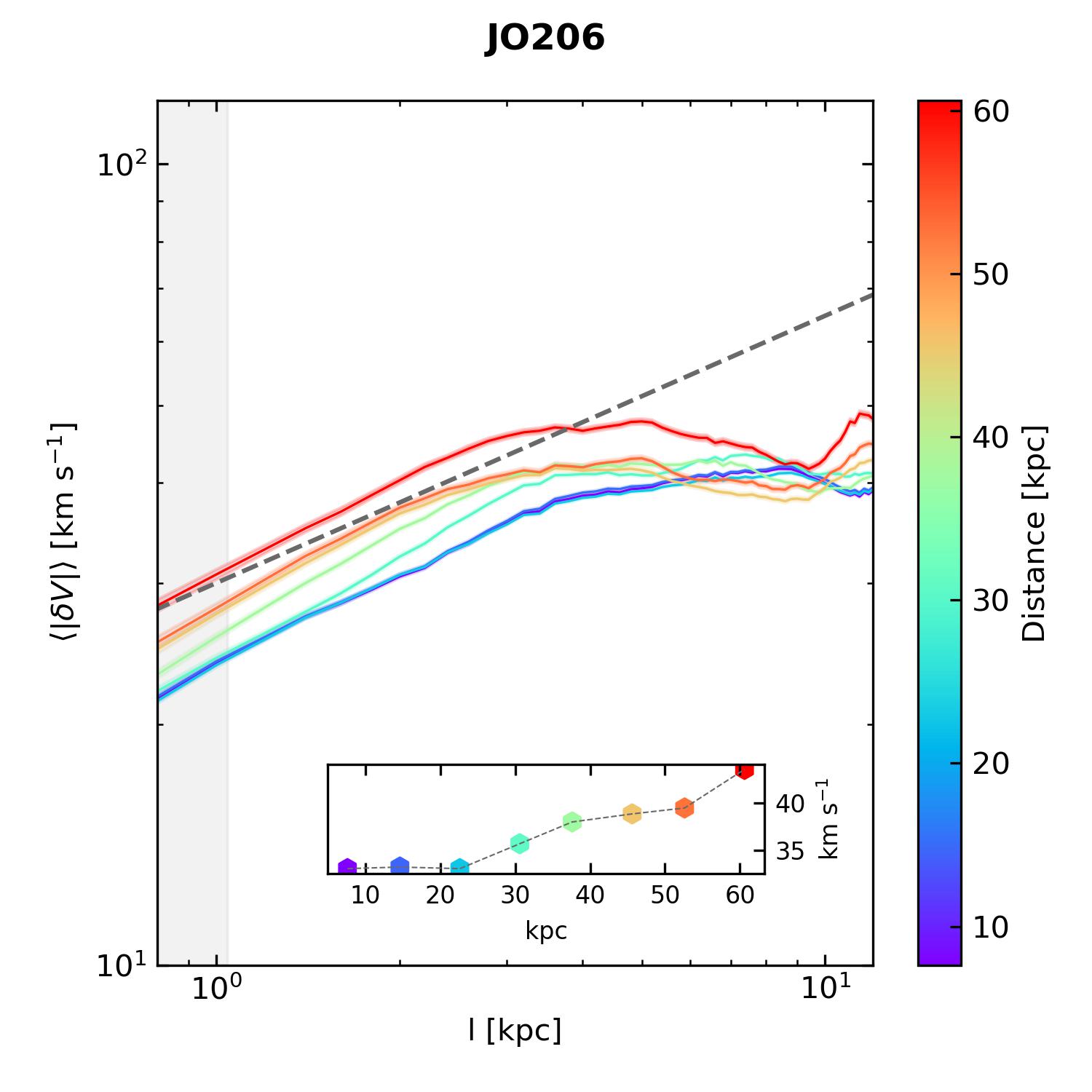}
\includegraphics[width=.5\textwidth]{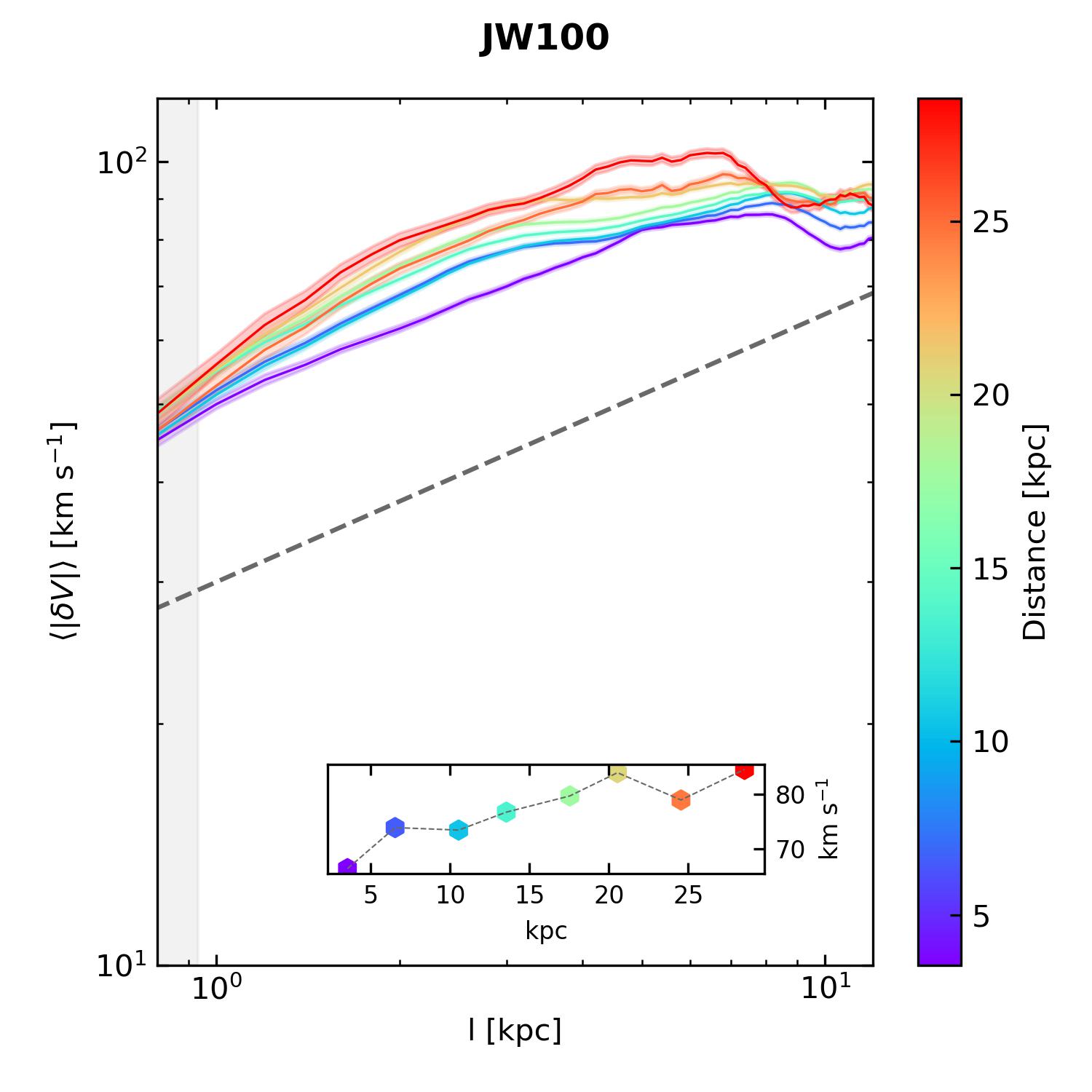}
    \caption{ \label{turbo_gradient} VSFs at increasing distance from the disk for JO147, JO204, JO206 and JW100. The VSFs are computed within a moving frame along the tail direction, color-coded by the distance of the frame center from the stellar disk edge. The dashed line indicates the Kolmogorov slope. In the inserts, we report the average VSF value between 2 and 3 kpc at increasing distances from the disk. }
   
\end{figure*}
To further understand how turbulence develops in the tails of GASP galaxies, we study the variation of the residual VSF at increasing distances from the stellar disk. We perform this analysis on JO147, JO204, JO206, and JW100, the galaxies with the longest tails in the sample. Similar to the approach presented by \citet[][]{Li_2023}, we define an adaptive moving frame, with a width corresponding to one-third of the tail length, in which we compute the VSF in eight bins at increasing distance from the stellar disk. We also compute the average value of each VSF between 2 and 3 kpc to quantify the trend with the distance. The results are shown in Figure \ref{turbo_gradient}.

The resulting VSFs broadly follow the $l^{1/3}$ slope below 5 kpc, as Figure \ref{img_VSF} shows. We observe a gradient with the distance, with the farthest bins showing a higher level of turbulence with respect to those close to the disk. This result is in line with a scenario where the mixing with the ICM is increasing the stripped ISM turbulence velocity with respect to the initial stellar disk conditions \citep[][]{Fraternali_2002,Iorio_2017,Bacchini_2020}. Therefore, the observed VSFs support our hypothesis that environmental turbulence ultimately drives the stripped gas small-scale motions.

\subsection{Implications for the ICM microphysics}
Our working hypothesis, based on the observational evidence of ISM-ICM mixing (see Section \ref{sec:intro}), is that the stripped ISM clouds can trace the local ICM turbulence. The increase in turbulence power with the distance from the disk and the fast dissipation of the small-scale turbulence with respect to the dynamical time, both discussed in Section \ref{3.2}, indicate that the turbulence observed the residual VSFs is generated outside of the disk, which is consistent with our assumption. Therefore, similarly to the case discussed by \citet[][]{Li_2023}, the observed, residual VSFs can provide us with new insights into the ICM properties. The ICM turbulent cascade is supposed to extend down to a dissipation scale beyond which the viscous forces dominate over the inertial ones, and thus the kinetic energy gets dissipated into heating. Therefore in the classical treatment of the ICM as a fluid driven by Coulomb collision, the smallest VSF scale would be the so-called Kolmogorov microscale $\eta$:
\begin{equation}
    \eta=\left(\frac{\nu^3}{\epsilon}\right)^{\frac{1}{4}}\text{ cm,}
    \label{eta}
\end{equation}
where $\epsilon$ is the specific energy flux, and $\nu$ is the kinematic viscosity. The latter is related to the dynamic viscosity, $\mu$, which can be computed as:
\begin{equation}
    \mu=\rho\nu=5500 \left(\frac{T_e}{10^8\text{ K}} \right)^{\frac{5}{2}} \left(\frac{\text{log}\Lambda}{40} \right)^{-1}\text{ g cm}^{-1}\text{s}^{-1},
    \label{visco_spitz}
\end{equation}
where log$\Lambda$ is the Coulomb logarithm \citep[][Equation 5.33]{Sarazin_1988}, and $\rho$ and $T_e$ are, respectively, the ICM mass density and electron temperature \citep[e.g.,][]{Spitzer_1962}. The specific energy flux $\epsilon$ can be computed as:
\begin{equation}
    \epsilon\simeq\frac{v^3}{l}\text{ cm}^{2}\text{s}^{-3}.    
\end{equation}

To compute $\eta$ it is thus necessary to know the local ICM electron density and temperature near the galaxies, which can be provided by the X-ray spectral analysis of the ICM thermal emission. We collect these values for eight out of ten galaxies, for which an X-ray spectral analysis is reported in the literature (see Table \ref{tab_kolmo}). For three of them, namely JO113, JO147, and JO162 only the ICM temperature is available, thus we compute a range for $\nu$ for the typical range of electron density $n_e=(0.1-1)\times10^{-3}$ cm$^{-3}$. Finally, for each galaxy, we estimate $\epsilon$ by using the VSF value at the reference scale of 3 kpc, that is the scale at which every galaxy shows a Kolmogorov-like slope. We note that for $v\propto l^{1/3}$, $\epsilon$ is scale-invariant. We estimate $\eta$ to be $<100$ kpc, which is in line with the previous estimates \citep[][]{Zhuravleva_2019,Li_2023}. For comparison, we also compute the electron mean free path $\lambda_e$ following \citet[][]{Spitzer_1956}:
\begin{equation}
    \lambda_e=\frac{3^{3/2}\left(kT_e\right)^2}{4\pi^{1/2}n_ee^4\text{log}\Lambda}\text{ cm,}
    \label{mfp}
\end{equation}
where $k$ is the Boltzmann constant, $T_e$ and $n_e$ are the ICM electron temperature and particle density, and $e$ is the electron charge. For JO113, JO147, and JO162 we compute the $\lambda_e$ range corresponding to $n_e=(0.1-1)\times10^{-3}$ cm$^{-3}$. We show the values of temperature, $kT_e$, and electron density, $n_e$, at the cluster-centric distance of each galaxy, VSF at 3 kpc, and the resulting $\eta$ and $\lambda_e$ in Table \ref{tab_kolmo}.

\begin{figure}
    \centering
     \includegraphics[width=.5\textwidth]{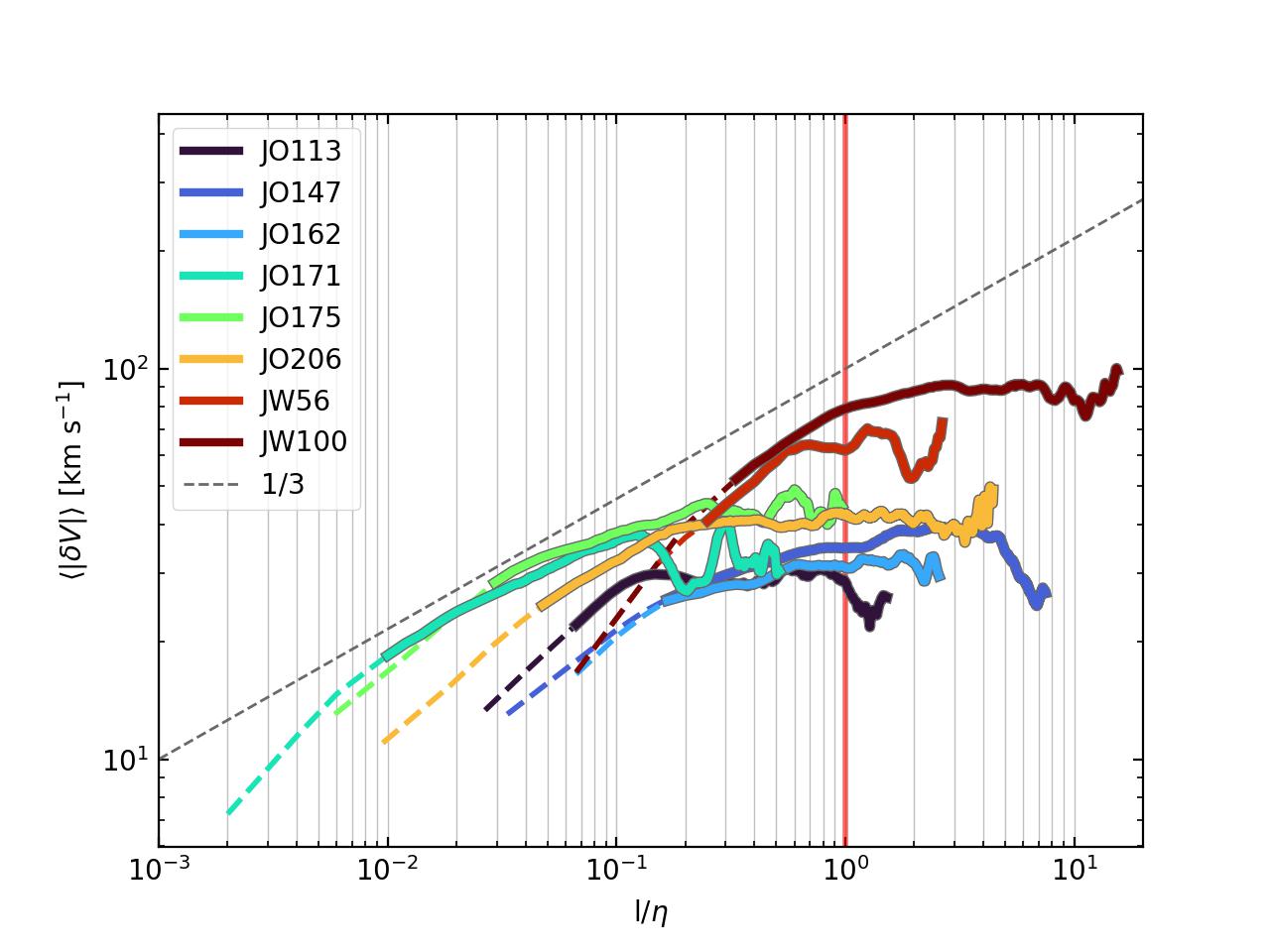}
    \caption{VSFs re-scaled to the corresponding $\eta$ reported in Table \ref{tab_kolmo}. Below $l=1$ kpc the re-scaled VSFs are reported with the dashed lines. JO113, JO147, and JO162 are re-scaled for the lower limit of $\eta$. The dashed, grey line indicates the $l^{1/3}$ Kolmogorov scaling, and the vertical, red line points out the $l/\eta=1$ scale.  }
    \label{img_vsf_kologorov}
\end{figure}

\begin{table*}[]
    \centering
    \begin{tabular}{lccccccc}
    \midrule
    \midrule
    Galaxy &$\langle |\delta V |\rangle_{\text{3 kpc}}$&$n_e$&$kT_e$&$\eta$&$\lambda_e$&log$_{10}(\nu/\nu_S)$&Reference\\
    &[km s$^{-1}$]&[$10^{-3}$ cm$^{-3}$]&[keV]&[kpc]&[kpc]&&\\
    \midrule
    JO113&50&-&4.0&[15, 86]&[4, 47]&[-2.6, -1.6]&\citet[][]{Whelan_2022}\\
    JO147&38&-&2.3&[6, 38]&[1, 16]&[-2.1, -1.0]&\citet[][]{Bardelli_1996}\\
    JO162&30&-&2.0&[6, 31]&[1, 12]&[-2.0, -1.0]&\citet[][]{Bardelli_2002}\\
    JO171&25&0.4&5.7&100&25&-2.7&\citet[][]{Akamatsu_2012}\\
    JO175&36&0.3&3.3&34&13&-2.0&\citet[][]{Andrade-Santos_2021}\\
    JO204&53&-&-&-&-&-\\
    JO206&36&0.5&3.2&22&6&-1.8&\citet[][]{Muller_2021}\\
    JW39&41&-&-&-&-&-\\
    JW56&63&2.0&2.8&4&1&-0.8&\citet[][]{Cavagnolo_2009}\\
    JW100&79&3.1&3.5&3&1&-0.6&\citet[][]{Poggianti_2019}\\
    
    \midrule
    \end{tabular}
    \caption{ \label{tab_kolmo}Results of the VSF analysis. From left to right: galaxy name; VSF at 3 kpc; ICM electron density and temperature at the galaxy cluster-centric distance; Kolmogorov micro-scale from Equation \ref{eta}; Particle mean free path from Equation \ref{mfp}; Ratio between the observed and predicted ICM viscosity; References for $n_e$ and $kT$. In square brackets, we give the range of values corresponding to $n_e=(0.1-1)\times10^{-3}$ cm$^{-3}$. }
   
\end{table*}
To test if the VSFs stop at the expected dissipation scale, we re-scale each VSF for the corresponding $\eta$, and, for JO113, JO147, and JO162, we use the lower limit of the interval of $\eta$ (Table \ref{tab_kolmo}). The resulting VSFs are presented in Figure \ref{img_vsf_kologorov}, in which we also show their components below $l=1$ kpc (dashed lines). It emerges that the VSFs easily extend below the expected dissipation scale, which corresponds to $l/\eta=1$ (red vertical line), with JO113, JO171, and JO175 completely developing in the regime $l/\eta<1$. This result is consistent with the previous study by \citet[][]{Li_2023}, and it confirms that the ICM dissipation scale is smaller than predicted by Equation \ref{eta}.

The observed and predicted ICM viscosity discrepancy, $\nu/\nu_S$, where $\nu_S$ is the Spitzer viscosity (Equation \ref{visco_spitz}), can be inferred by comparing the observed dissipation scale, $l_d$, with the predicted one, $\eta$. Specifically, from Equation \ref{eta} follows that:
\begin{equation}
    \frac{\nu}{\nu_S}=\left(\frac{l_d}{\eta}\right)^{\frac{4}{3}}\text{ .}
\end{equation}
We note that in our work we cannot reliably sample the VSF below $\sim1$ kpc due to the atmospheric seeing (see Section \ref{sec_data}), thus we conservatively estimate the upper limit for $\nu/\nu_S$ by setting $l_{d}=1$ kpc. We report the resulting log$_{10}(\nu/\nu_S)$ in Table \ref{tab_kolmo}. Our results indicate that the real ICM viscosity is smaller than the Spitzer estimate, with a difference that spans, at least, two orders of magnitude from log$_{10}(\nu/\nu_S)=-0.6$ (JW100) to log$_{10}(\nu/\nu_S)=-2.7$ (JO171), in line with the previous results \citep[][]{Zhuravleva_2019,Li_2023}. 

This discrepancy could be evidence that the ICM particles' mean-free path is smaller than the predicted $\lambda_e$ (Table \ref{tab_kolmo}) due to plasma effects. Since we are measuring the VSFs at scales that are similar to, or smaller than, the Coulomb scale $\lambda_e$, for completeness we note that it is also possible that self-organization effects in collisionless turbulence \citep[e.g., magneto-immutability,][]{Squire_2019} may allow forming a magneto hydro dynamic-like turbulent inertial range also below the putative viscous scale \citep[][]{Majeski_2024}. In this case, rather than a reduced effective $\lambda_e$, we are probing effects due to collisionless physics in the ICM turbulence. We also note that the viscosity can be further suppressed of a factor $\sim3$ by magnetic fields \citep[e.g.,][]{Mogavero_2014, Komarov_2018}. The presence of magnetic fields in the jellyfish galaxies's tails is demonstrated by the existence of nonthermal radio continuum emission, which constrains the typical intensity to be between 4 and 7 $\mu$G \citep[e.g.,][]{Chen_2020,Ignesti_2021,Ignesti_2022d,Roberts_2021a,Roberts_2021b,Ignesti_2023}.

\subsection{Implications for the stripped ISM}
The presence of turbulence in the stripped ISM and the detection of a minimum scale can have implications for the local star formation \citep[e.g.,][]{MacLow_2004, Elmegreen_2014,Federrath_2021} because below the dissipation scale, turbulence does not significantly contribute to the gas internal pressure that contrasts the gravitational collapse. For JO175, JO204, JO206, JW39, and JW100 \citet[][]{Giunchi_2023} report the size distribution of star-forming clumps, finding that those in the tails of these galaxies are indeed limited to sizes of $0.3$ kpc, that would be in agreement with the dissipation scales derived in our work. Remarkably, the complexes in the tails result smaller than those in the stellar disk, hinting at different, underlying processes shaping the origin of star-forming complexes in these different environments. 

Additionally, in Figure \ref{img_VSF} we observe that the VSF, measured at the reference value of 3 kpc, ranges from $\sim30$ to 90 km s$^{-1}$. This gradient could be due to the work impressed by the ram pressure during the galaxy orbit, which can be tentatively outlined by the position of these galaxies in the phase-space diagram. In this diagram, which we show in Figure \ref{img_phase_space}, galaxies in the top-left corner show the highest velocity and the smallest clustercentric distance, thus they should be subject to the strongest ram pressure. We note however that the phase-space coordinates are only lower limits of the corresponding quantities. JW100 and JW56, which show the highest line-of-sight velocities and the smallest projected distances, also show the highest $\langle |\delta V |\rangle_{\text{3 kpc}}$ values, namely 79 and 63 km s$^{-1}$ (Table \ref{tab_kolmo}). Therefore, the data tentatively support the hypothesis of a connection between ram pressure and VSF amplitude, but a larger sample is required to confirm it.

\section{Caveats}
\label{sec_caveats}
\begin{enumerate}
\item The procedure to mitigate the effect of rotation and advection is based on the distance of each pixel from the stellar disk. Due to the complex morphology of these galaxies, this measure is not univocal. Different methods may result in a different filament geometry in the phase-space plots, especially near the stellar disk, which may affect the outcomes of this procedure;
\item This analysis is subject to projection effects. The stripped tails are composed of filaments of ionized plasma, and we cannot discriminate if two spaxels in two filaments that appear nearby on the plane of the sky are physically close, or if they are in distant locations in the tail. Therefore, we can only measure a lower limit of the true physical separation. The projection effect can result in a bias at any given scale contaminated by large velocity differences at larger scales, resulting in a global increment in the observed turbulence power. Furthermore, previous studies suggest that projection effects can also slightly flatten the VSF, so that the true 3D VSF may be steeper than the projected, 2D case \citep[][]{Ganguly_2023, Chen_2023}. However, we note that the simulated tail analysis (Figure \ref{fig_test_VSF}) suggests that the different projections can only increase the VSF amplitude at small scales without changing its slope (W45 vs. W45-y). We also note that the velocity pairs at the edge of adjacent filaments are less than those inside each filament, thus their potentially biased contribution is subdominant in the computation of the VSF; 
\item The estimate of the dissipation scale $\eta$ is based on the observed ICM conditions, derived from X-ray spectroscopy, in the galaxies' surroundings. However, X-ray observations of jellyfish galaxies \citep[e.g.,][]{Machacek_2011,Zhang2013,Poggianti_2019,Campitiello_2021,Bartolini_2022,Sun_2022} have shown that these galaxies can also host hot, X-ray-emitting plasma in their tails, with typical temperatures between 0.8 and 1.2 keV. This hot component is believed to be the hottest part of the multi-temperature mixing layer forming in the contact surface between ISM and ICM, thus it is reasonable to expect it to play a part in driving the turbulence in the ISM. Due to the lower temperature, the dissipation scale in this mixing layer is likely smaller than in the ICM (see Figure \ref{param-K}), and, potentially, it could reach scales of $\eta\simeq1$ kpc. This scale is consistent with our upper limit and larger than the 0.2 kpc constrained in \citet[][]{Li_2023} (black dashed line). Therefore, even assuming that the turbulence is driven solely by the mixing layer does not rule out the tension between expected and observed viscosity. Nevertheless, the filling factor of the stripped filaments is unknown, thus we cannot exclude the possibility that the turbulence development inside them is regulated by other factors with significant local variability which the average ICM properties may not well represent; 
    \begin{figure}
        \centering
        \includegraphics[width=\linewidth]{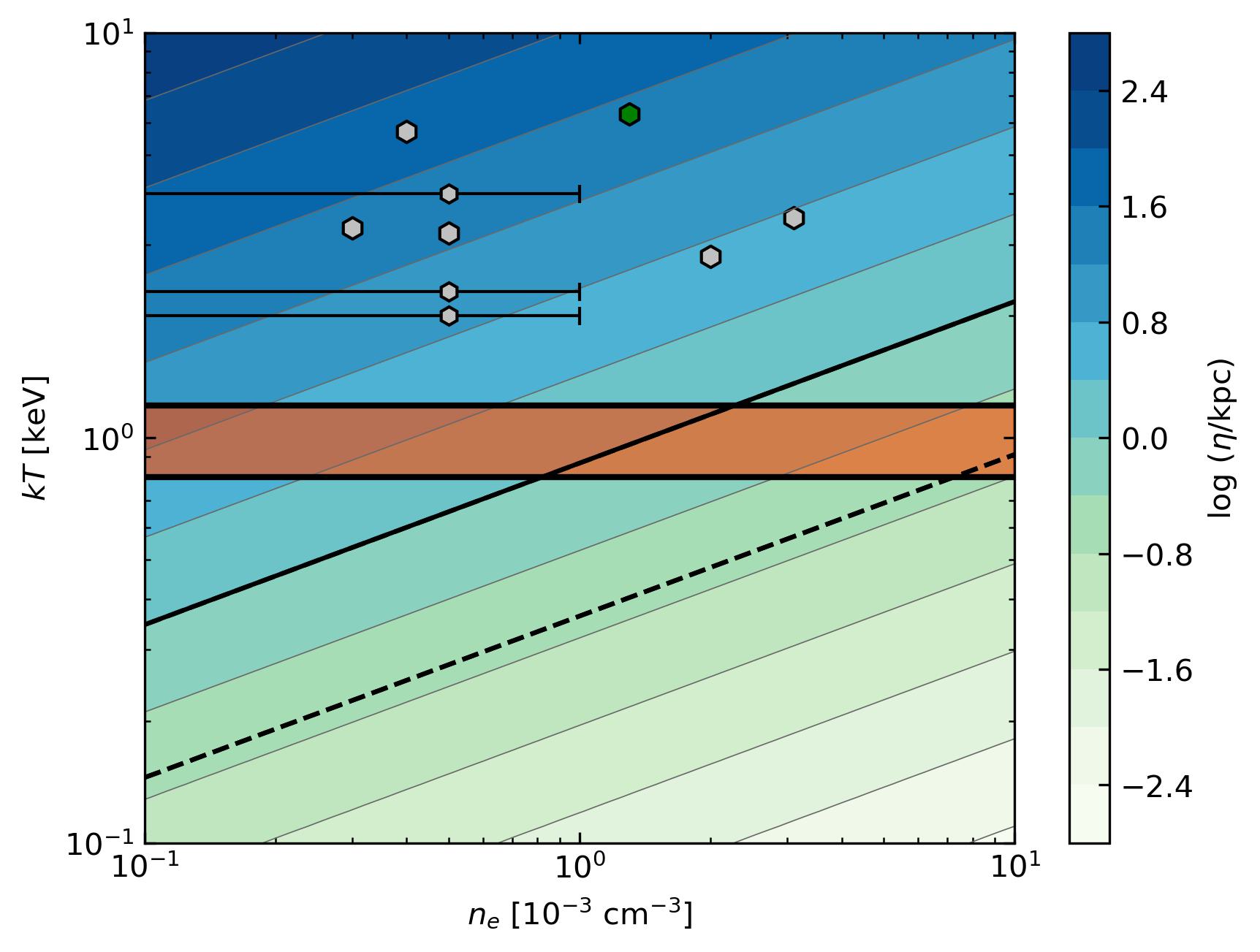}
        \caption{\label{param-K} Parameter space of $\eta$, computed according to Equation \ref{eta}, for $n_e=10^{-4}-10^{-2}$ cm$^{-3}$, $kT=0.1-10$ keV, and reference values $l=3$ kpc and $\langle |\delta V |\rangle_{\text{3 kpc}}=44$ km s$^{-1}$, which is the average value measured in this work. We report the ICM conditions of the cluster sample (grey points, see Table \ref{tab_kolmo}) and the one reported in \citet[][]{Li_2023} (green point), the $\eta=1$ kpc and $\eta=0.2$ kpc (black continuous and dashed lines), and the typical temperature range of the mixing layers, 0.8-1.2 keV (red band).}
    \end{figure}
\item We are assuming that the injection scale of the turbulence is significantly larger than the dissipation scale, but we cannot fully exclude that the observed VSFs might result from multiple injections at different scales. However, the comparison between dissipation and dynamical time scales presented in Section \ref{3.2} indicates that turbulence injected below the scale of a few kpc in the disk is fully dissipated in the clouds at the end of the tails. So we argue that, in case of multiple injections, recreating the observed VSFs would require a continuous injection of small-scale turbulence into the stripped ISM at increasing distances from the disk, which is unlikely. 
\end{enumerate}

\section{Summary and conclusions}
In this paper, we study the development of turbulence in the tails of ten GASP jellyfish galaxies by reconstructing the VSF of the H$\alpha$ emission observed with MUSE. Our working hypothesis, supported by observational evidence, is that the VSF of the diffuse plasma in the stripped tails could trace the turbulent motions of the surrounding ICM, with which the ISM clouds are mixing. Therefore, the extent and the shape of the ISM VSF can constrain the ICM properties, especially its viscosity. This is especially interesting because the ICM viscosity is currently poorly constrained, and there is a tension between the predictions based on Coulomb collision theory, and the observations.

The observed VSFs resemble the slope $\propto l^{1/3}$ from 1 kpc, the smallest scale reliably sampled by our observations, up to a few kpc. By comparing our results with those obtained by measuring the VSF on simulated tails with different orientations, we confirm that a combination of rotation and advection cannot solely explain the measured slope. The simulated tail study suggests that the effect of projection effects is limited to slightly increasing the VSF amplitude without affecting its slope. We also observe that the velocity increases with the distance from the stellar disks. So we argue that our results are consistent with the hypothesis that the stripped ISM small-scale motions, at least in the terminal part of the tails, are driven by the surrounding ICM turbulence.

Then, we use previous X-ray studies in the literature to compute the expected ICM viscosity and the corresponding dissipation scale, and we compare them with the observed VSFs. It clearly emerges that the VSFs extend for a couple of orders of magnitudes below the supposed dissipation scale set by the ICM. This result sets the actual ICM viscosity at 0.3-25$\%$ of the expected value.

The implications of our results are manifold. We have presented a new piece of evidence that, on small scales, it is necessary to account for the effect of plasma instabilities to describe the ICM microphysics. The low viscosity implies that either the particles have a smaller mean free path than expected in a regime defined by Coulomb collisions, or the ICM turbulence is probing collisionless plasma effects. Concerning the case of jellyfish galaxies, our results indicate that the tails are turbulent down to sub-kpc scales, which can locally influence the process of gas collapse and star formation.
\section*{Acknowledgments}
We thank the Referee for the constructive report that improved the presentation of this work. We thank S. Tonnesen for the useful discussions. This project has received funding from the European Research Council (ERC) under the European Union's Horizon 2020 research and innovation programme (grant agreement No. 833824, PI Poggianti) and the INAF founding program 'Ricerca Fondamentale 2022' (PI A. Ignesti). C.B. acknowledges support from the Carlsberg Foundation Fellowship Programme by Carlsbergfondet. Y.L. acknowledges financial support from NSF grants AST-2107735 and AST-2219686, NASA grant 80NSSC22K0668, and Chandra X-ray Observatory grant TM3-24005X. 

Based on observations collected at the European Organization for Astronomical Research in the Southern Hemisphere under ESO programme 196.B-0578.  This research made use of Astropy, a community-developed core Python package for Astronomy \citep[][]{astropy_2013, astropy_2018}, and APLpy, an open-source plotting package for Python \citep[][]{Robitaille_2012}. A.I. thanks the music of Fearless Flyers for inspiring the preparation of the draft.

\bibliography{sample631}{}
\bibliographystyle{aasjournal}
\appendix

\section{Data preparation I: Phase-space plots}
\label{app_rot_corr}
\begin{figure}
    \includegraphics[width=.5\textwidth]{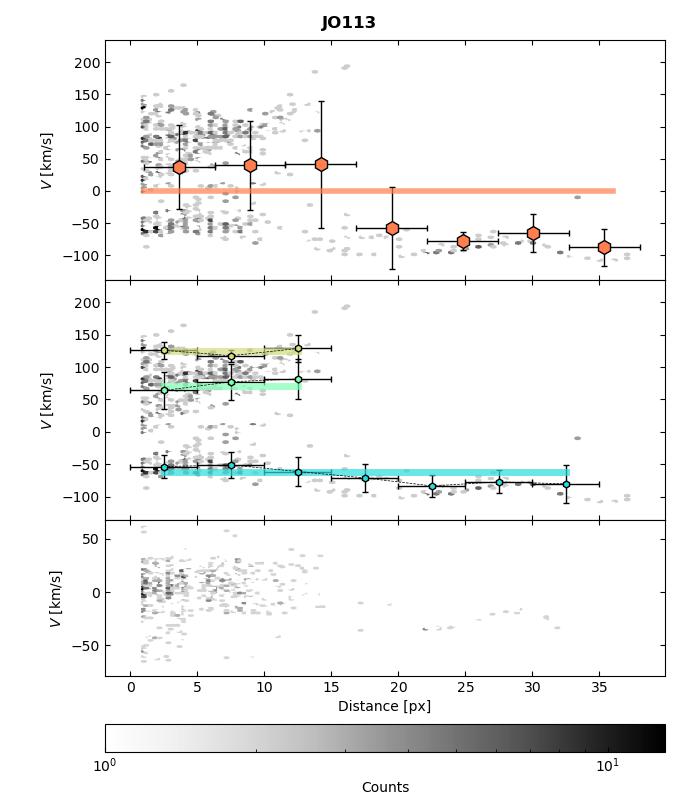}
    \includegraphics[width=.5\textwidth]{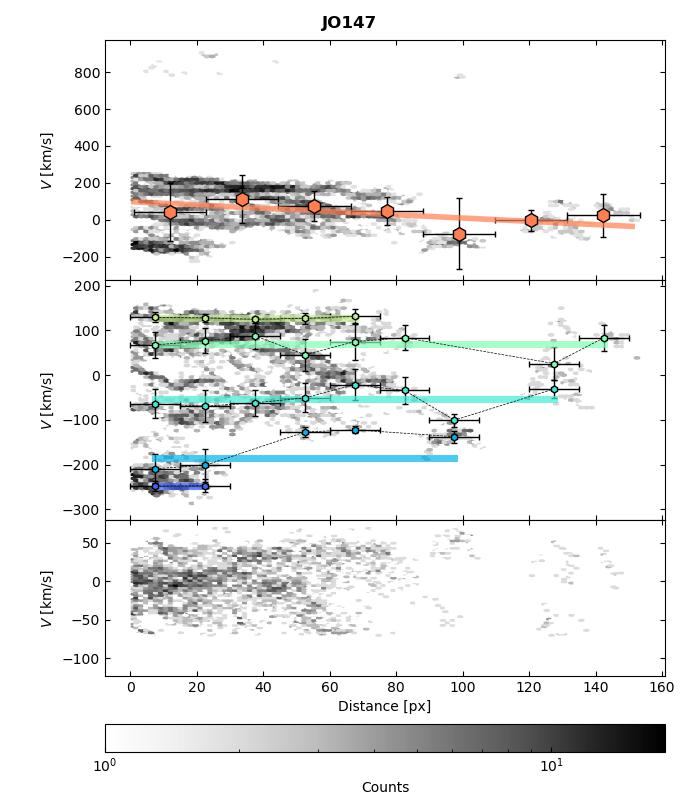}\\
    \includegraphics[width=.5\textwidth]{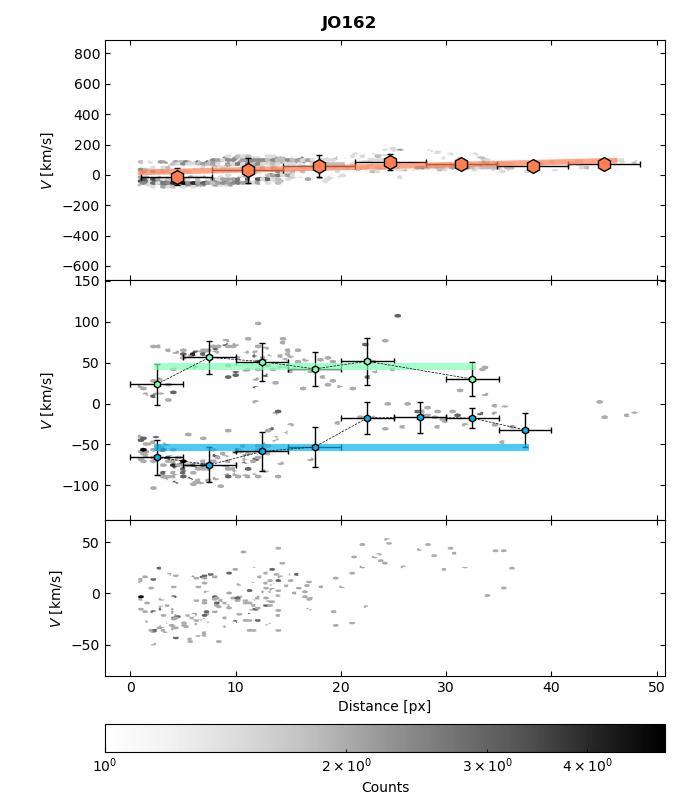}
    \includegraphics[width=.5\textwidth]{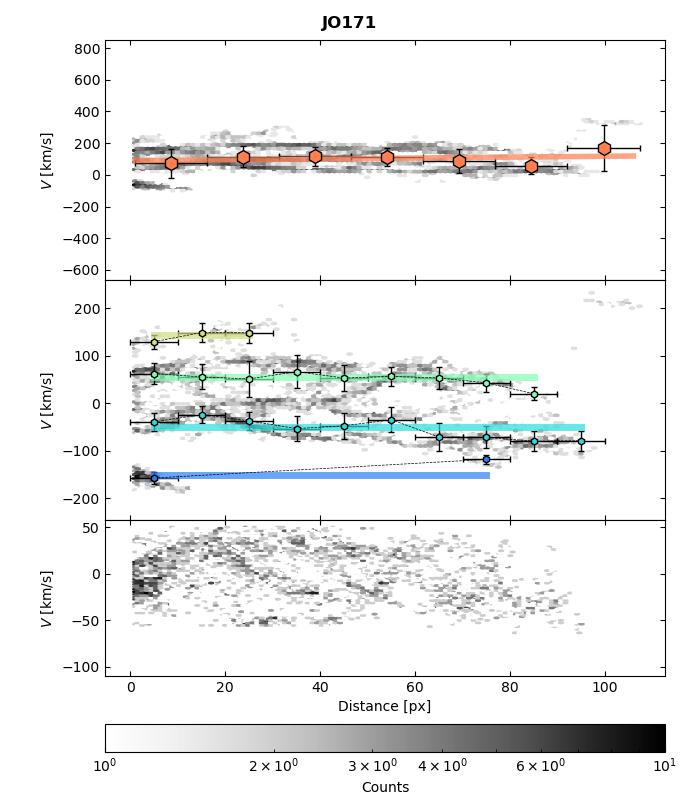}\\
    \caption{\label{app_fig1}Phase-space density plots for JO113, JO147, JO162, and JO171.}
\end{figure}

We report in Figure \ref{app_fig1} the phase-space density plots for each galaxy except JO206, which is shown in Figure \ref{vel_grad}. Top: the initial state with the best-fit slope modeling the systematic advection velocity (orange line); Middle: phase-space density plot after the correction for the systematic velocity, overlapped by the results of the filament finder; Bottom: Phase-space density plot of the residual velocity map.
\renewcommand{\thefigure}{\arabic{figure} (Cont.)}
\addtocounter{figure}{-1}
\begin{figure}
    \includegraphics[width=.5\textwidth]{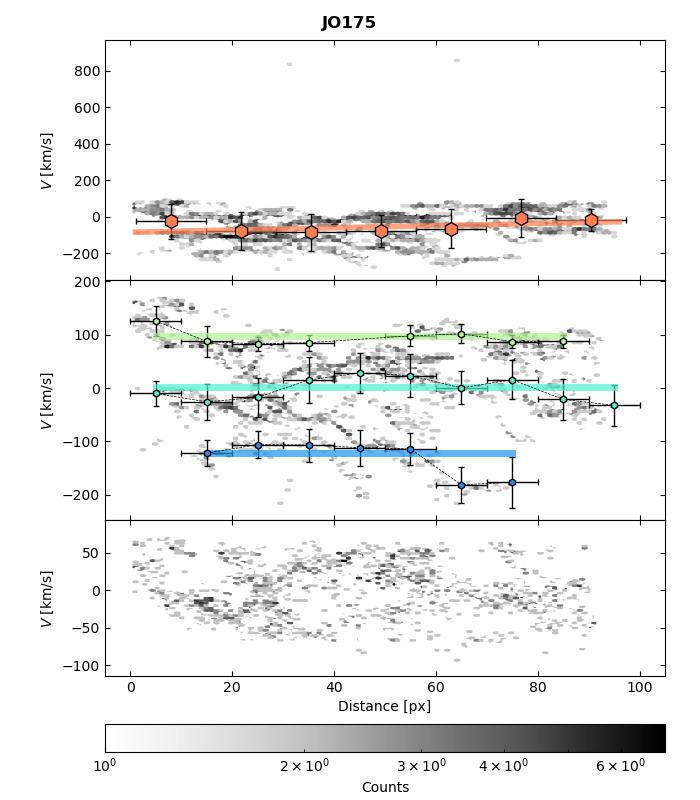}
    \includegraphics[width=.5\textwidth]{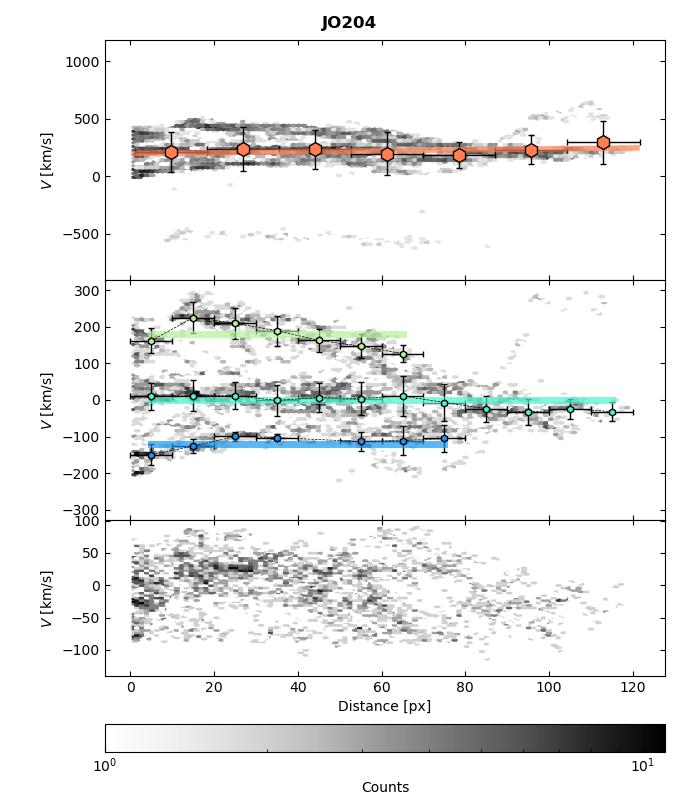}\\
    \includegraphics[width=.5\textwidth]{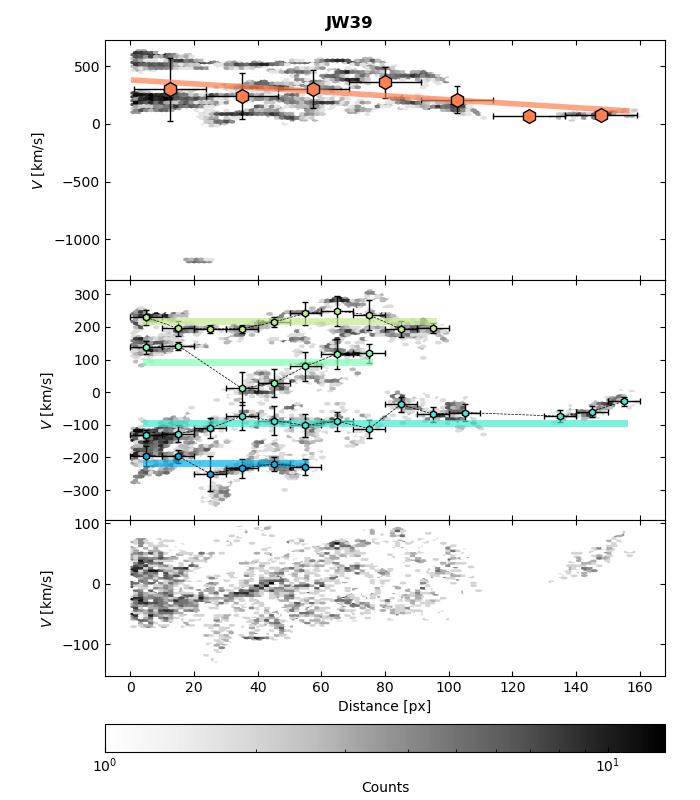}
    \includegraphics[width=.5\textwidth]{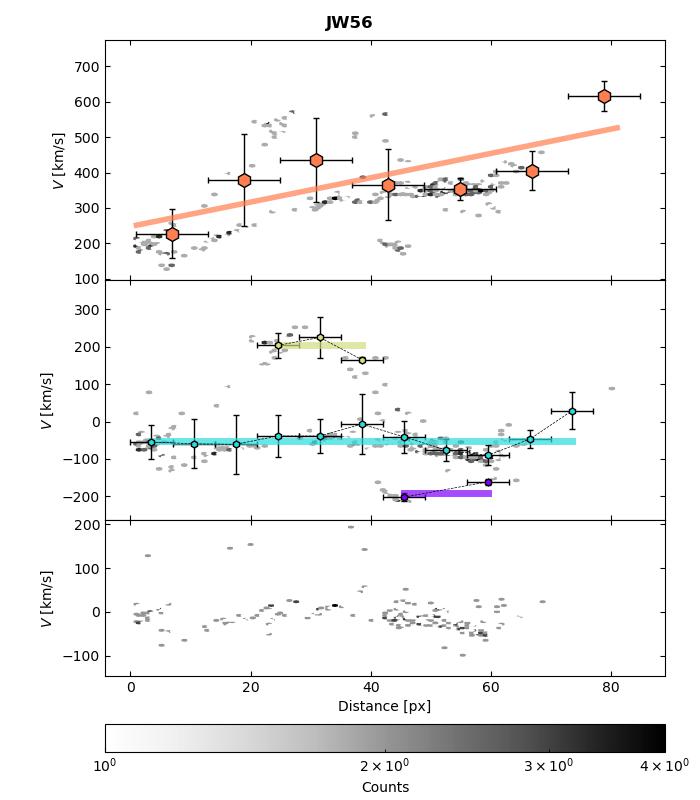}\\
    \caption{Phase-space density plots for JO175, JO204, JW39, and JW56.}
\end{figure}
\renewcommand{\thefigure}{\arabic{figure} (Cont.)}
\addtocounter{figure}{-1}
\begin{figure}
    \centering
    \includegraphics[width=.5\textwidth]{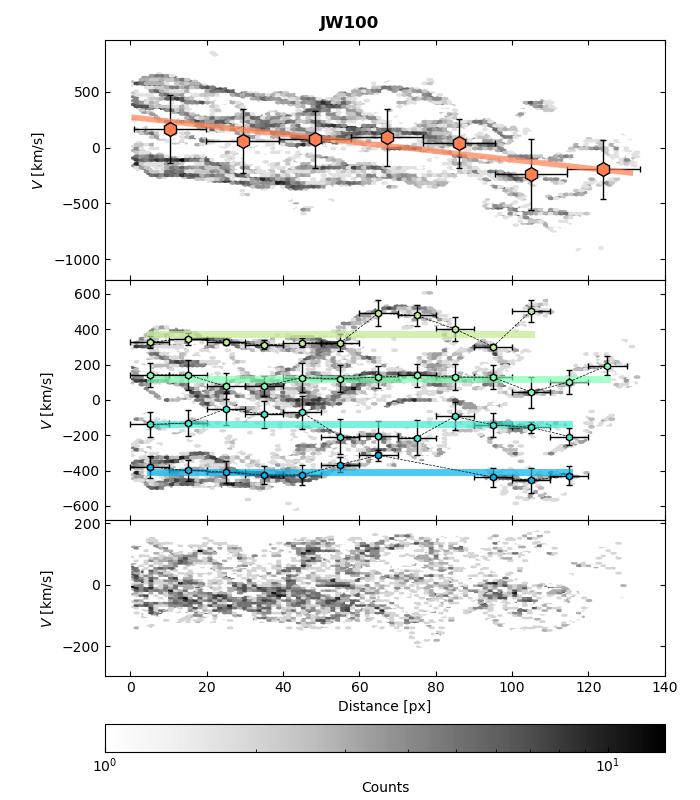}
    \caption{Phase-space density plots for JW100}
\end{figure}
\renewcommand{\thefigure}{\arabic{figure}}

\section{Data preparation II: Models}
\label{app_models}
\begin{figure}
    \includegraphics[width=.32\textwidth]{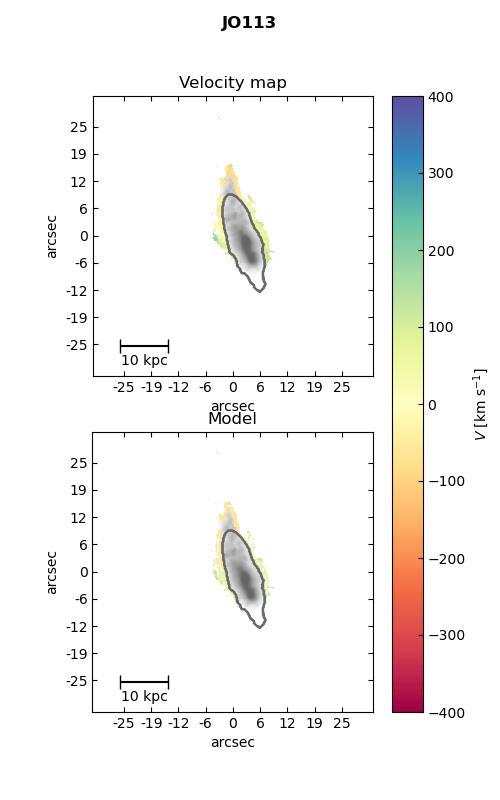}
    \includegraphics[width=.32\textwidth]{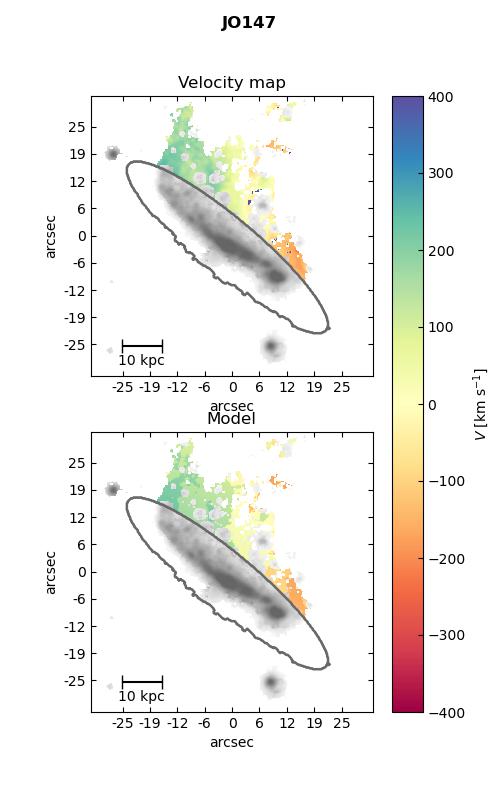}
    \includegraphics[width=.32\textwidth]{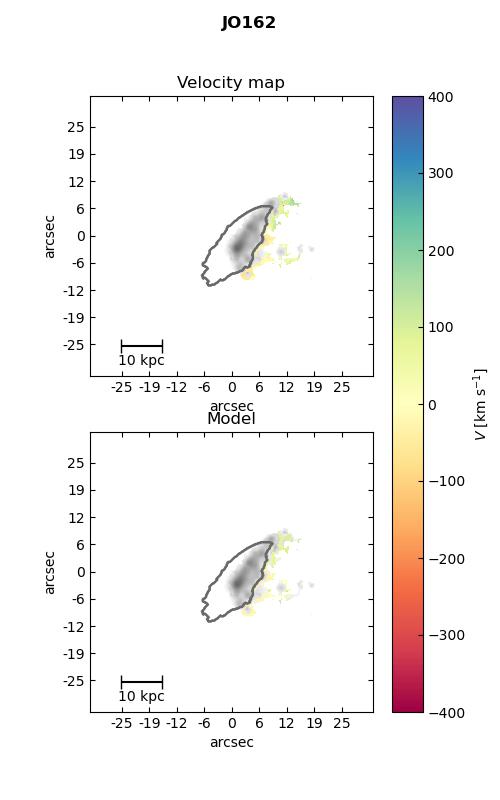}\\
    \includegraphics[width=.32\textwidth]{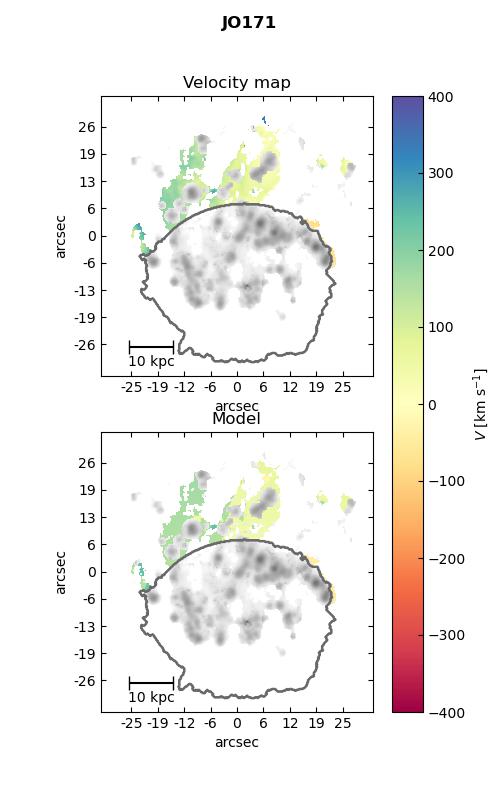}
    \includegraphics[width=.32\textwidth]{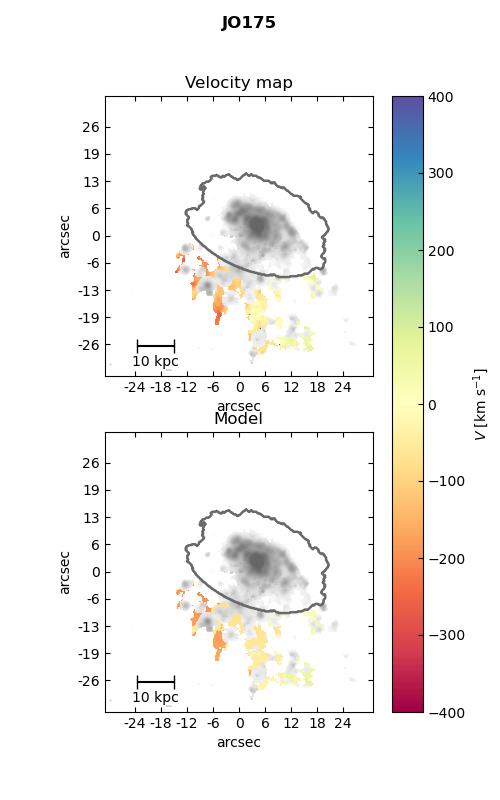}
    \includegraphics[width=.32\textwidth]{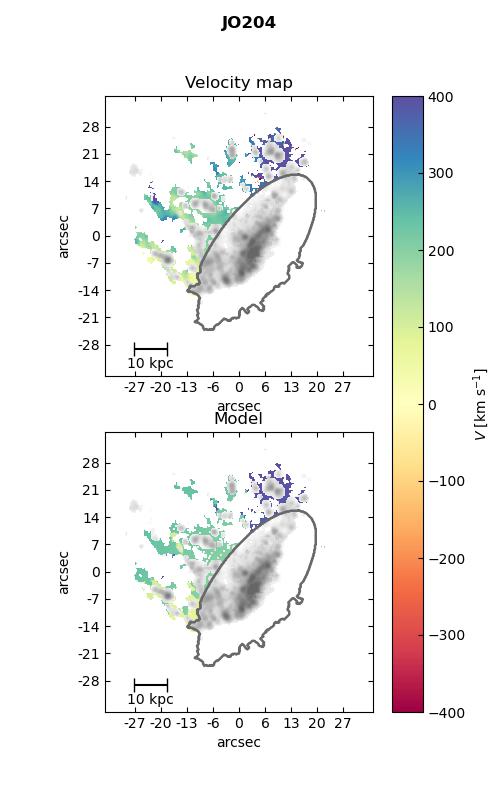}
    \caption{\label{app_fig2}Observed (top) vs modeled (bottom) H$\alpha$ velocity maps for JO113, JO147, JO162, JO171, JO175, and JO204.}
\end{figure}

In Figure \ref{app_fig2} we report the comparison between the observed H$\alpha$ velocity maps (top) and the corresponding models (bottom). For each galaxy, we also show the original H$\alpha$ emission (grey colormap), and the stellar disk \citep[grey contour][]{Gullieuszik_2020}.
\renewcommand{\thefigure}{\arabic{figure} (Cont.)}
\addtocounter{figure}{-1}
\begin{figure}
    \includegraphics[width=.32\textwidth]{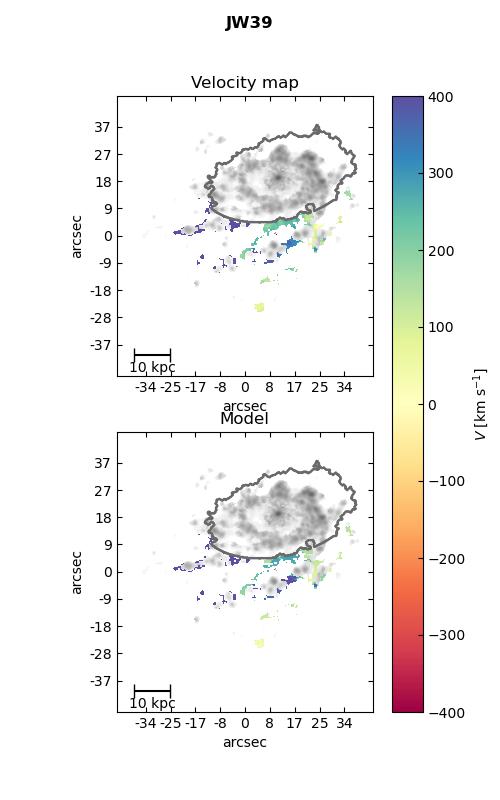}
    \includegraphics[width=.32\textwidth]{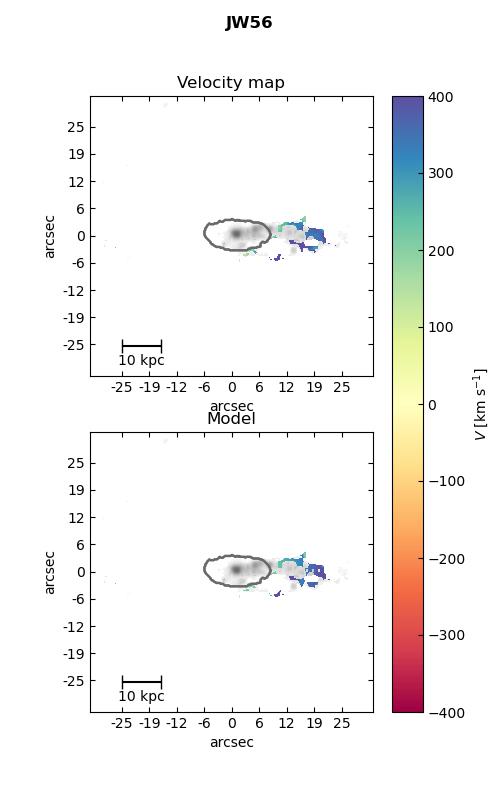}
    \includegraphics[width=.32\textwidth]{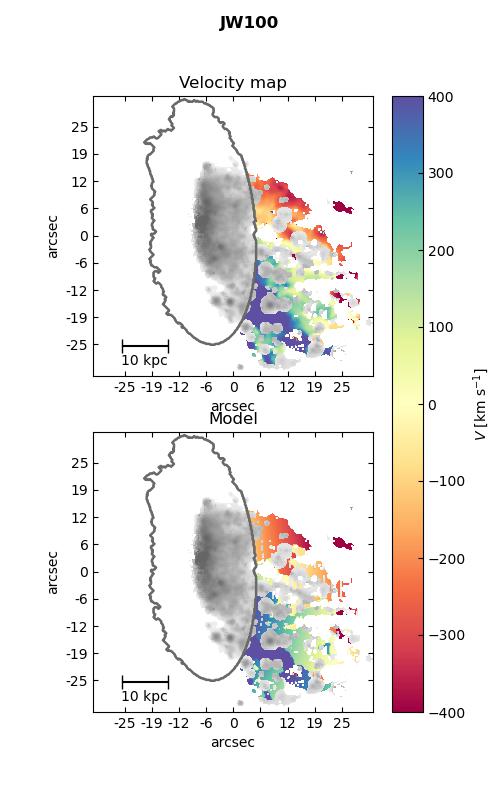}
    \caption{Observed (top) vs modeled (bottom) H$\alpha$ velocity maps for JW39, JW56, and JW100.}
\end{figure}
\renewcommand{\thefigure}{\arabic{figure}}

\section{Sample VSF}
\label{appendix_vsf}
\begin{figure*}[htb]
\centering
\includegraphics[width=.9\linewidth]{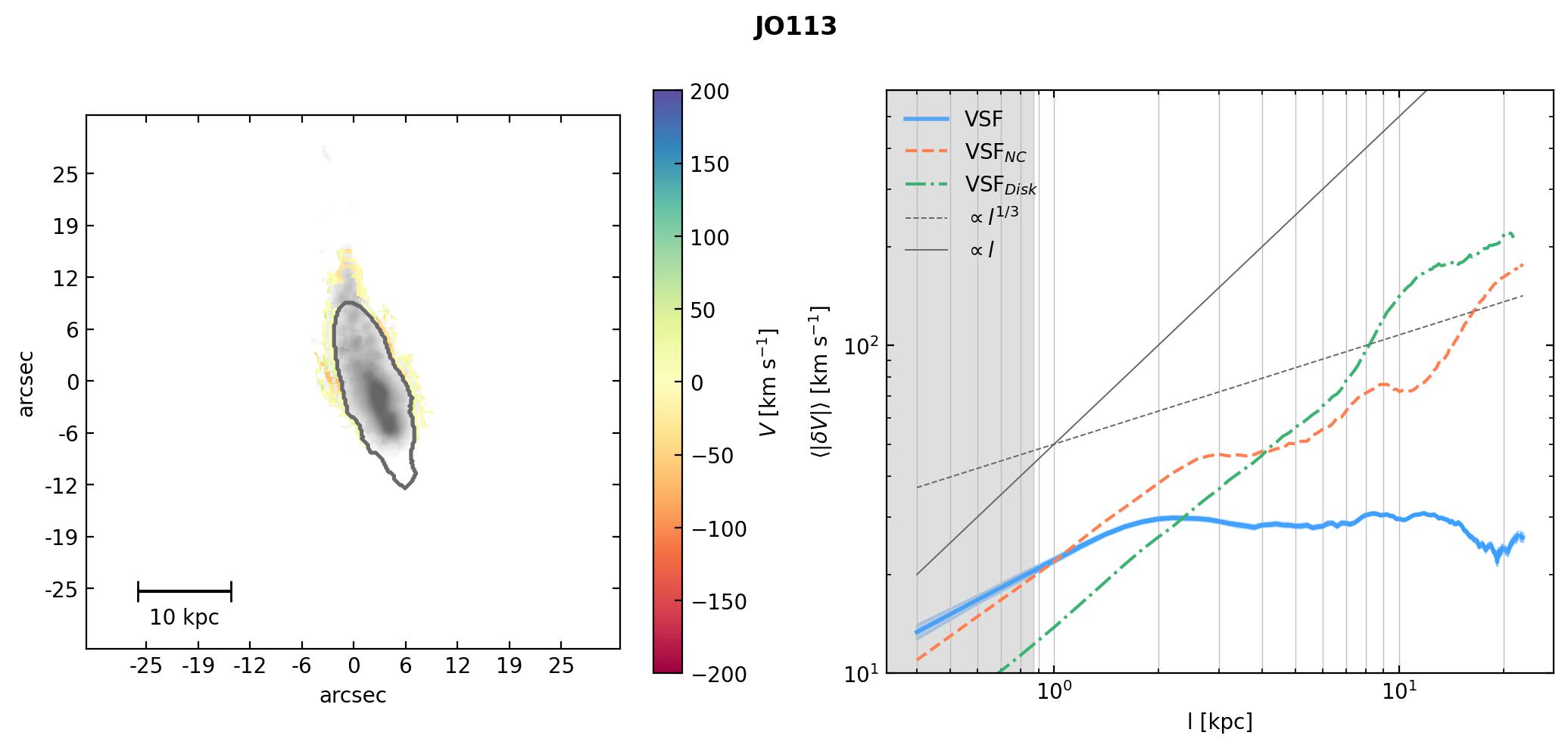}
\includegraphics[width=.9\linewidth]{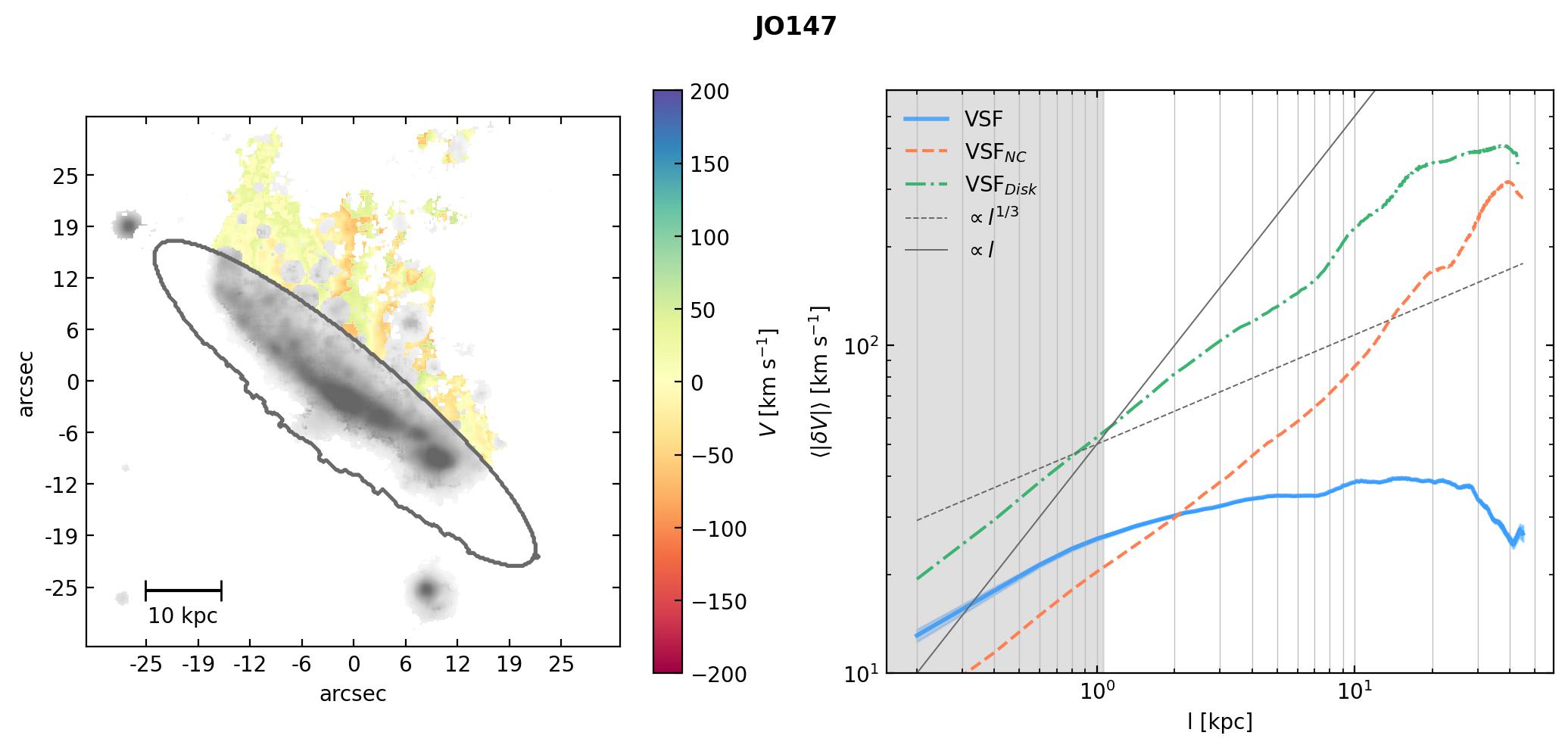}
\includegraphics[width=.9\linewidth]{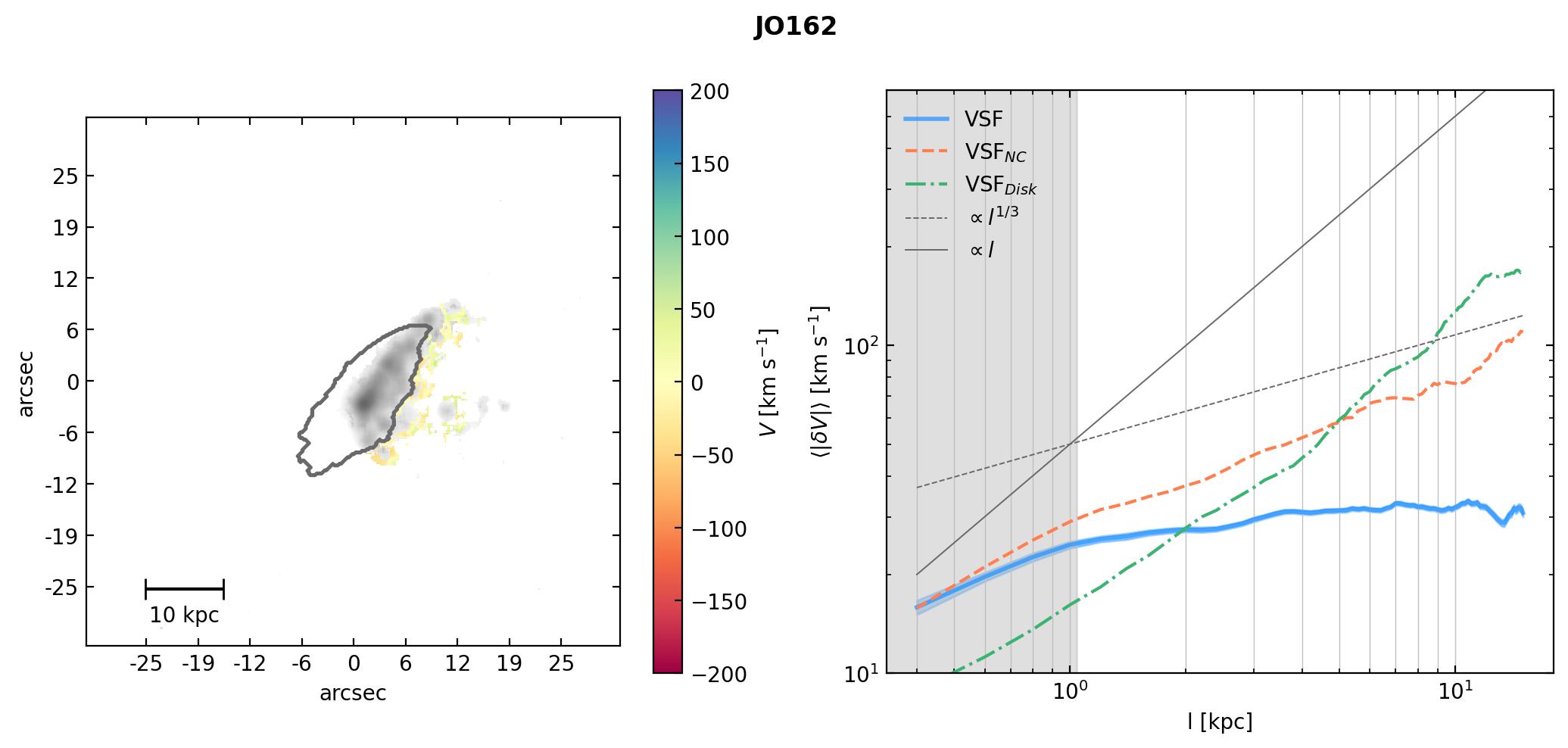}
\caption{\label{app_vsf}Residual H$\alpha$ velocity map and VSF for JO113 (top), JO147 (middle), and JO162 (bottom).}
\end{figure*}

We report in Figure \ref{app_vsf} the H$\alpha$ velocity maps and the corresponding VSF for each sample galaxy. In the left panels, we show the residual H$\alpha$ velocity map composed of the spaxels used to compute the VSF (blue-to-red colormap), the original H$\alpha$ emission (grey colormap), and the stellar disk \citep[grey contour][]{Gullieuszik_2020}. In the right panels, we report the resulting VSF (blue, the blue-filled area indicates the 1-$\sigma$ uncertainty region), and the VSF computed before the correction (orange). The grey-shaded area marks $l<1$ arcsec below which the signal is affected by the spaxel correlation (See Section \ref{data_analysis}).

\renewcommand{\thefigure}{\arabic{figure} (Cont.)}
\addtocounter{figure}{-1}
\begin{figure*}[htb]
\centering

\includegraphics[width=.9\linewidth]{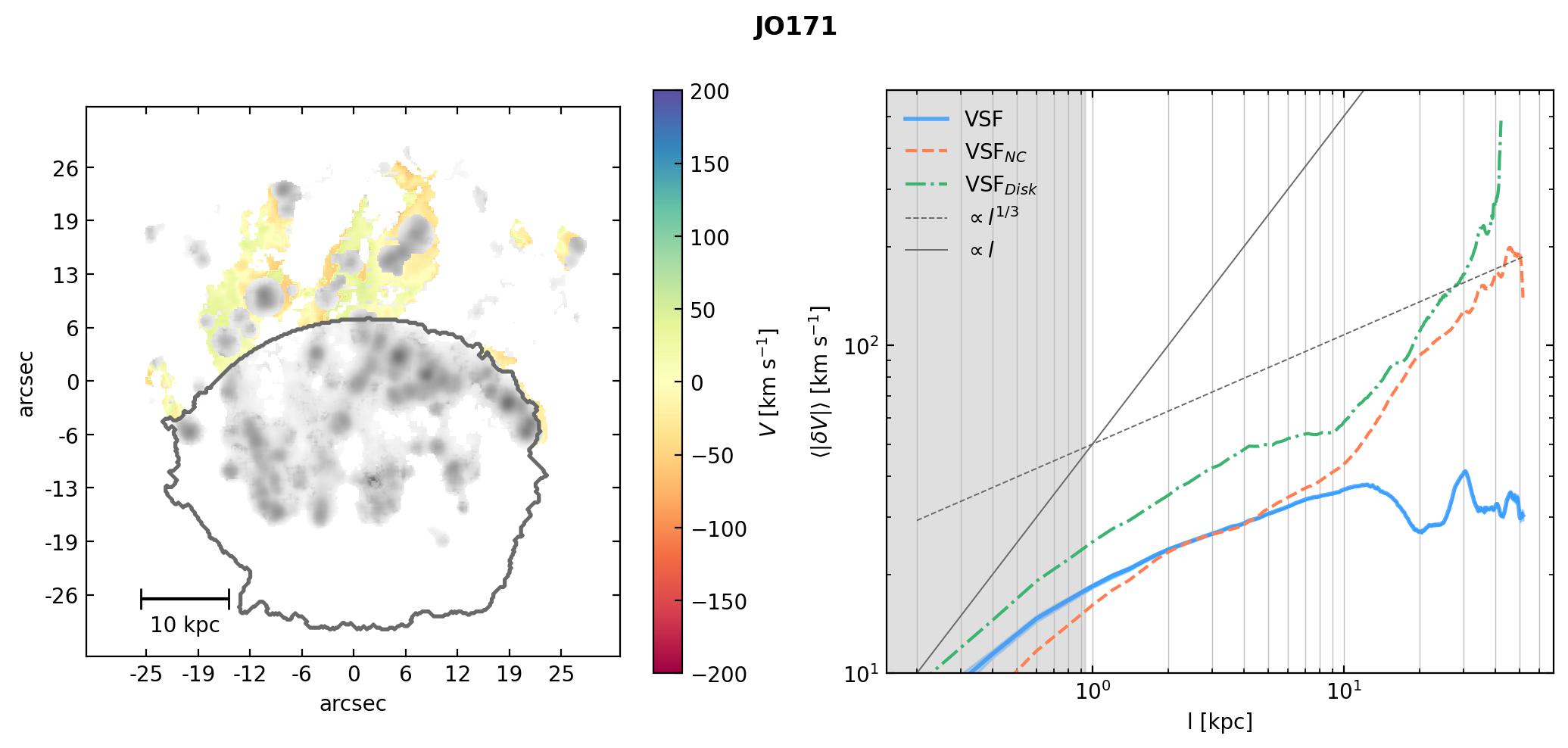}
\includegraphics[width=.9\linewidth]{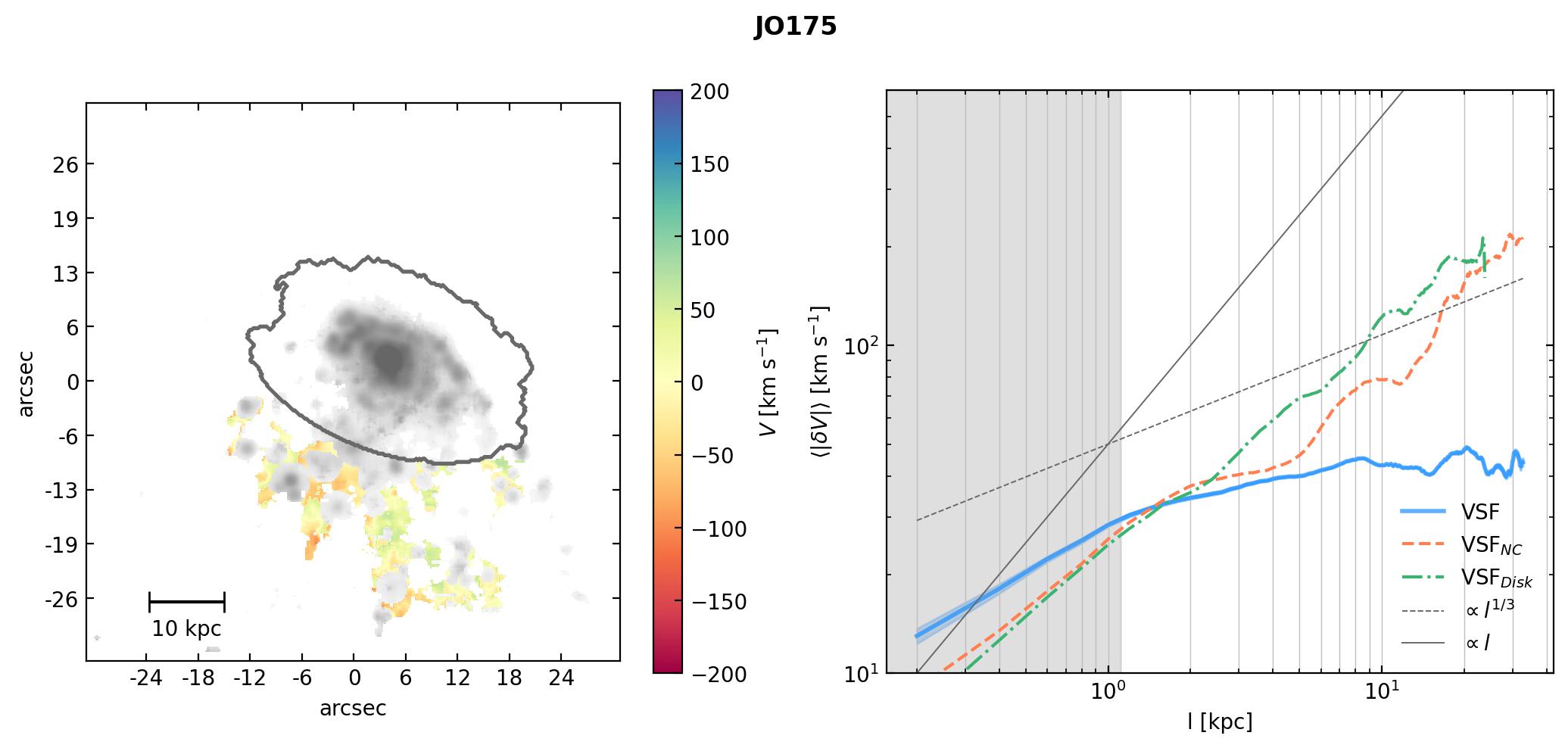}
\includegraphics[width=.9\linewidth]{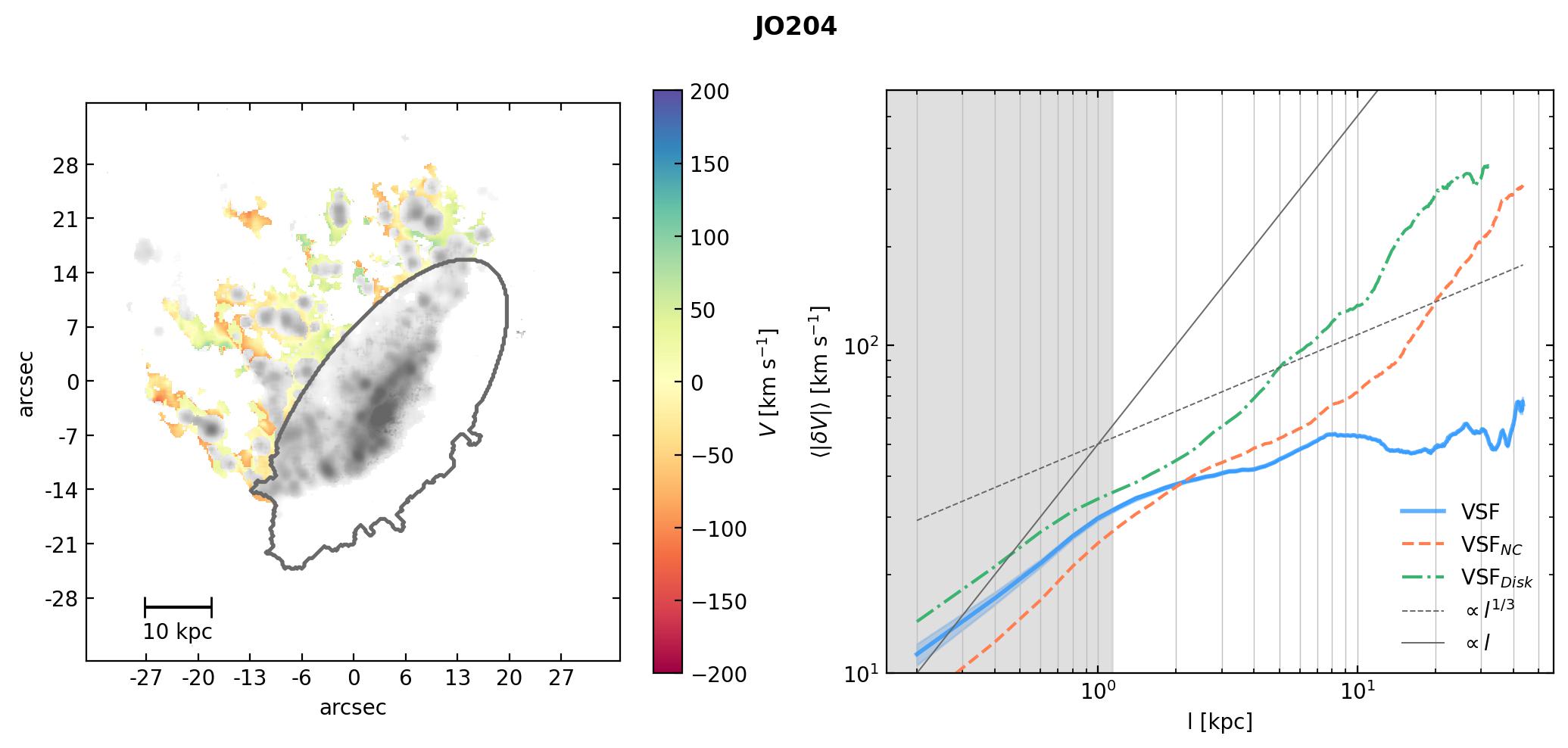}
\caption{Residual H$\alpha$ velocity map and VSF for JO171 (top), JO175 (middle), and JO204 (bottom).}
\end{figure*}
\renewcommand{\thefigure}{\arabic{figure} (Cont.)}
\addtocounter{figure}{-1}
\begin{figure*}[htb]
\centering

\includegraphics[width=.9\linewidth]{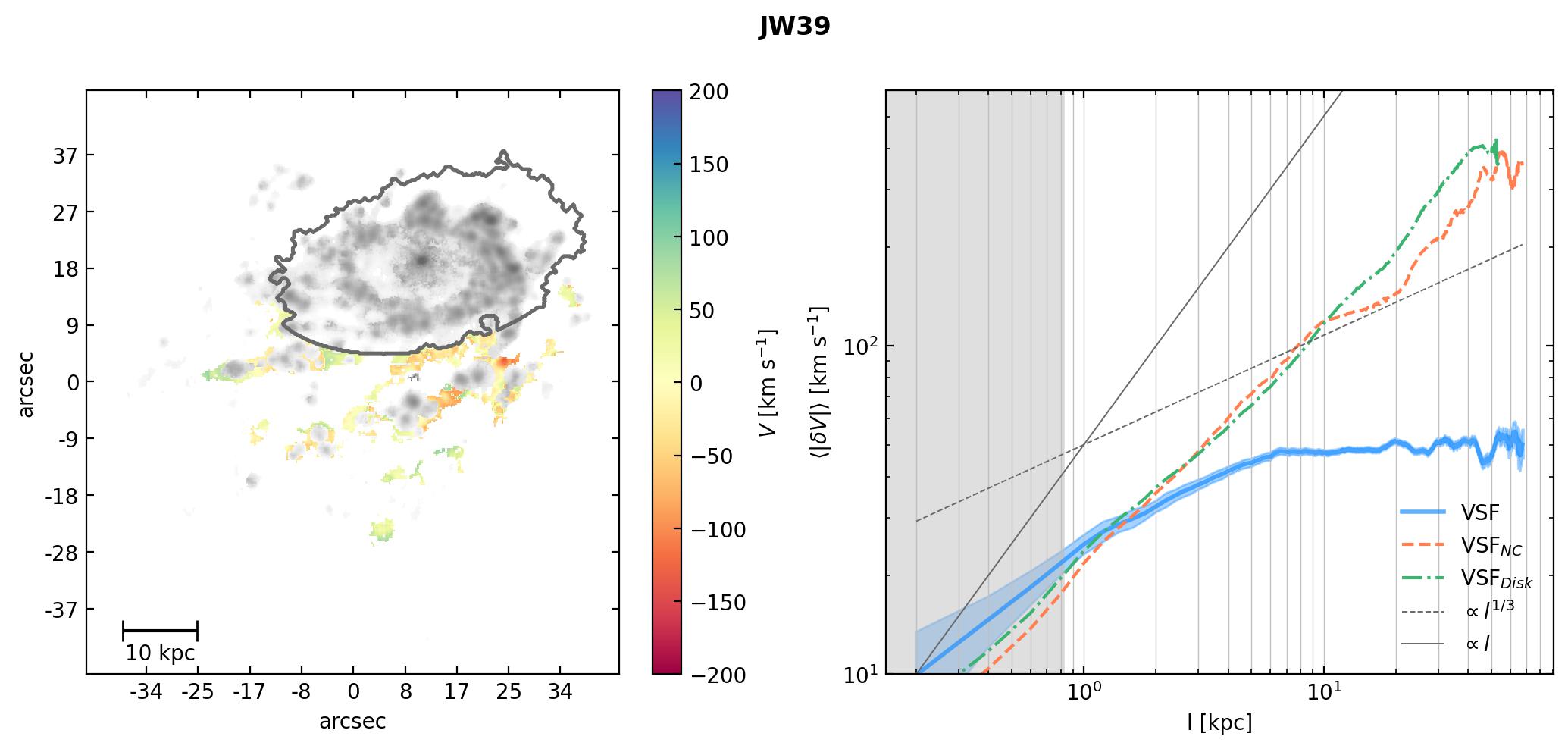}
\includegraphics[width=.9\linewidth]{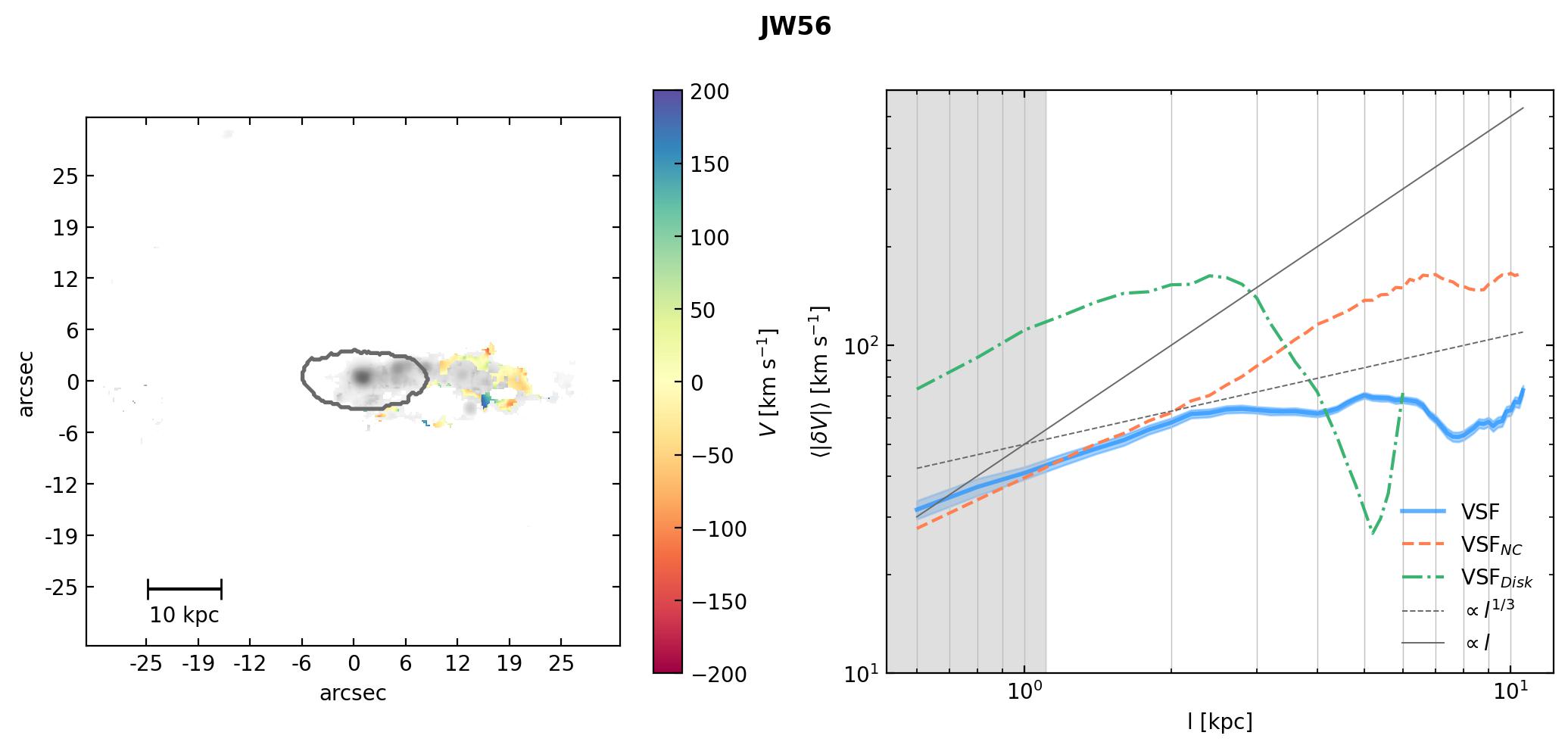}
\includegraphics[width=.9\linewidth]{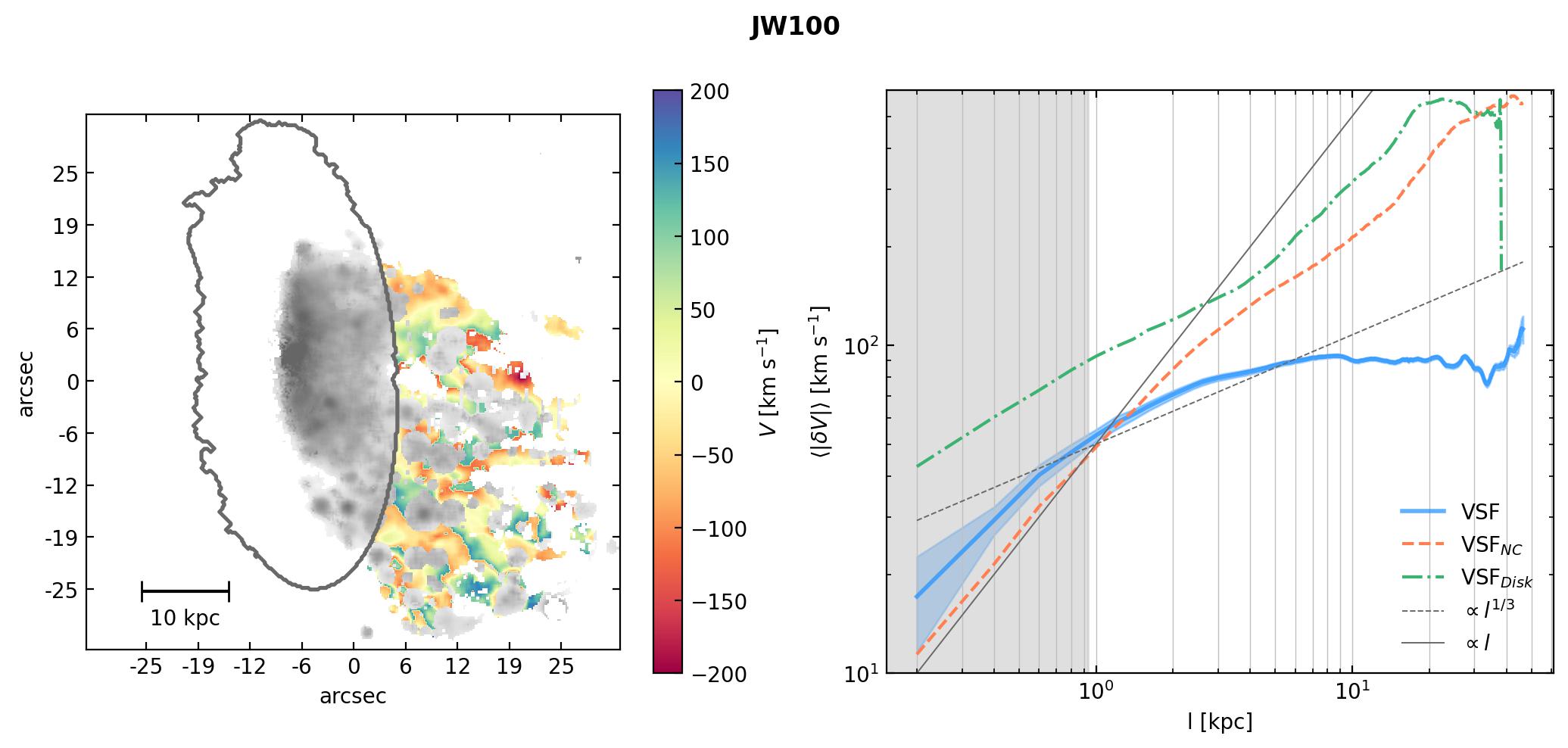}

\caption{Residual H$\alpha$ velocity map and VSF for JW39 (top), JW56 (middle), and JW100 (bottom).}
\end{figure*}
\renewcommand{\thefigure}{\arabic{figure}}
\section{Simulation set-up and images} \label{simu}

Here we briefly describe the simulation set-up presented in \cite{Akerman_2023}. The simulations were carried out using adaptive mesh refinement code Enzo \citep{Enzo}. The whole simulation box has 160 kpc on a side, and the cells can be refined according to the Jeans length and cell mass up to 39 pc. Radiative cooling is calculated using the {\sc grackle} library \citep{Grackle} and star formation and stellar feedback are modeled as in \cite{Goldbaum15, Goldbaum16}. A stellar particle of a mass of $1000 M_\odot$ will be formed if the density of a cell exceeds $10 \text{cm}^{-3}$ and assuming 1\% efficiency. Stellar feedback includes ionizing radiation from young stars, winds from evolved massive stars, and momentum input from supernovae.

The galaxy set-up follows \cite{RoedigerBruggen06} in which stellar disc and dark matter halo are represented by static potentials while the self-gravity of the gas is calculated at each time step. The Plummer–Kuzmin stellar disc \citep{MiyamotoNagai1975} has a mass of $10^{11} M_\odot$, scale length of 5.94 kpc and scale height of 0.58 kpc. Dark matter halo is modeled with a Burkert profile \citep{Burkert1995, MoriBurkert00} and has a radius of 17.36 kpc. The gaseous disc starts with a mass of $10^{10} M_\odot$, scale length of 10.1 kpc and scale height of 0.97 kpc.

The galaxy sits in the centre of the simulation box. To simulate RPS, we allow for the ICM wind to enter the box from one side (and outflow from the other). Ram pressure varies in strength, as the ICM velocity and density get bigger with time, imitating a galaxy on its first infall into a cluster. To calculate the wind parameters we follow the procedure described in \cite{Bellhouse2019} and assume a massive cluster of $10^{15} M_\odot$ and a beta model for the ICM. ICM is in hydrostatic equilibrium with gas temperature of $7.55\times10^7$ K. The galaxy is assumed to have started falling into the cluster from 1.9 Mpc clustercentric distance and with an initial velocity of 1785 $\text{km s}^{-1}$. The pre-wind ICM parameters are defined through Rankine–Hugoniot jump conditions for Mach number of 3 for the initial parameters of the ICM wind.

We perform two simulations in which the wind hits the galaxy at two different angles: face-on ($0^\circ$, wind flows along the $z$-axis) and $45^\circ$ (wind flows along both the $z$- and the $y$-axes). We refer to these galaxies as W0 and W45, respectively. The wind angle is kept constant throughout each simulation. The two galaxies evolve in isolation for 300 Myr before the wind reaches them and the stripping begins. During this time, the disc settles down and the variation in SFR goes down from 300\% to 5\% on a 5-Myr timescale.

\begin{figure}
    \centering
    \includegraphics[width=.8\linewidth]{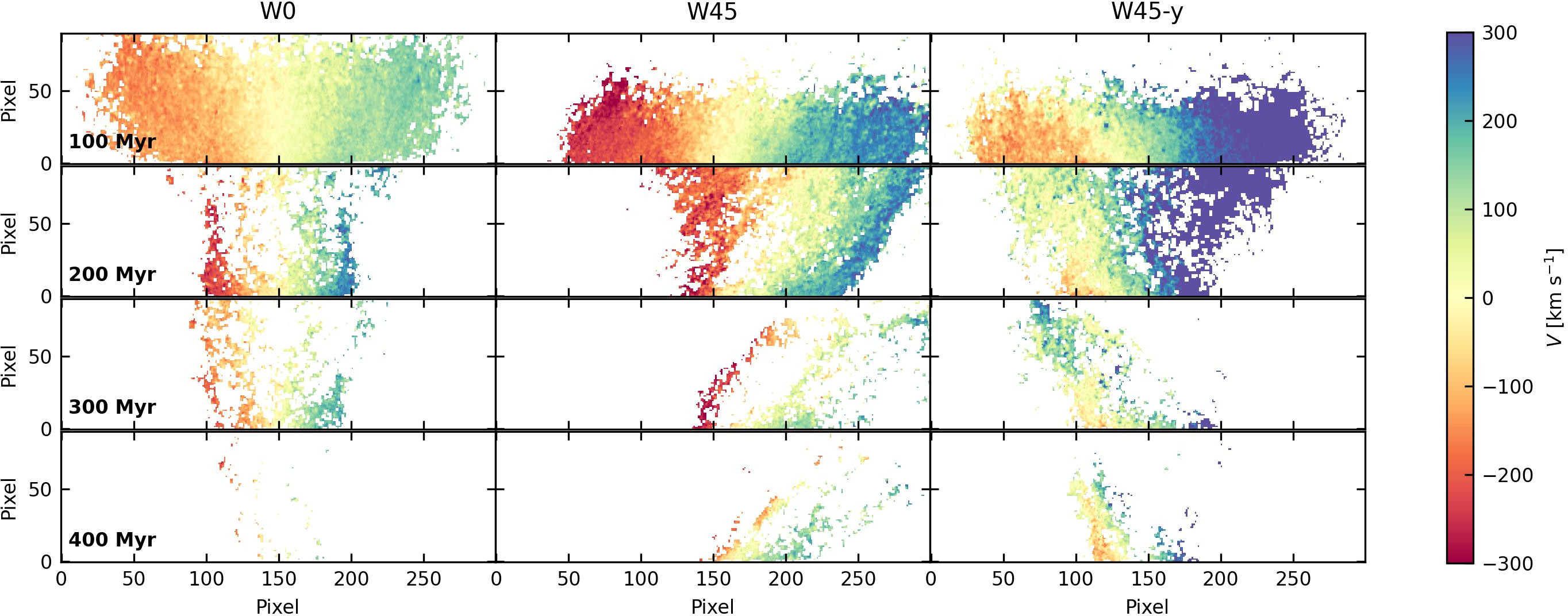}
    \caption{Velocity maps of the simulated tail, divided by simulation runs (from left to right, W0, W45, and W45-y) and epoch (from top to bottom, 100, 200, 300, and 400 Myr after the beginning of the stripping).}
    \label{im_simu}
\end{figure}

In Figure \ref{im_simu} we present the velocity maps for the three configurations. The gas has a temperature range of $3.5<$logT$<6.5$. In each panel, the sub-panels show, from top to bottom, maps for 100, 200, 300, and 400 Myr after the beginning of the stripping. The maps capture the gas in the stripped tails (defined as 2 kpc above the galaxy plane), and each pixel is 200 pc by 200 pc. The physical size of each map is 60 kpc by 18 kpc (up to 20 kpc from the galaxy plane).
We take two projections: one along the $x$-axis (left and middle columns for W0 and W45, respectively) and one along the $y$-axis (right column, `W45-y'). The latter allows us to study the projection effects on the velocity maps in W45, as we align the line-of-sight with the wind direction. 

The Figure shows how the amount of gas in the tails gradually decreases with time as the galaxy gets continuously stripped. The maps illustrate that the gas retains coherent rotation even while being stripped as far as 20 kpc from the galaxy plane and even at later phases of stripping. In W45-y, the gas has higher velocities as it gets an additional `kick' from the ICM wind. Here, at later times (300 and 400 Myr), the tail is directed to the left and not simply along the direction of stripping (which would be along the line of sight in this case). This is the result of the combined action of the wind and the galaxy rotation (see \cite{Akerman_2023} for an $x-y$ projection).

\end{document}